\documentclass[oneside]{ar-1col_hacked}
\usepackage{url}
\usepackage[numbers]{natbib}
\usepackage{amsmath,amssymb}
\usepackage{hyperref}
\usepackage[framemethod=TikZ]{mdframed}
\usepackage{amsthm}
\usepackage{multirow}
\usepackage{rotating}
\usepackage{soul}

\jname{}
\jvol{}
\jyear{}
\doi{}

\definecolor{cerulean}{rgb}{0.0, 0.48, 0.65}

\newcounter{lore}[section] 
\renewcommand{\thelore}{\arabic{section}.\arabic{lore}}
\newenvironment{loreyes}[2][]{%
\refstepcounter{lore}%
\mdfsetup{%
frametitle={%
\tikz[baseline=(current bounding box.east),outer sep=0pt]
\node[anchor=east,rectangle,fill=green!70!red!90]
{\strut Lore~\thelore};}}
\mdfsetup{%
singleextra={%
\node[anchor=east,rectangle,fill=green!70!red!90] at (11.5,0) {\strut #1};}}
\mdfsetup{innertopmargin=5pt,innerbottommargin=10pt,skipbelow=3pt,linecolor=green!70!red!90,%
linewidth=2pt,topline=true,%
frametitleaboveskip=\dimexpr-\ht\strutbox\relax
}
\begin{mdframed}[roundcorner=5pt,nobreak=true]\relax%
\label{#2}}{\end{mdframed}}
\renewcommand{\thelore}{\arabic{section}.\arabic{lore}}
\newenvironment{loreno}[2][]{%
\refstepcounter{lore}%
\mdfsetup{%
frametitle={%
\tikz[baseline=(current bounding box.east),outer sep=0pt]
\node[anchor=east,rectangle,fill=red!50]
{\strut Lore~\thelore};}}
\mdfsetup{%
singleextra={%
\node[anchor=east,rectangle,fill=red!50] at (11.5,0) {\strut #1};}}
\mdfsetup{innertopmargin=5pt,innerbottommargin=10pt,skipbelow=3pt,linecolor=red!50,%
linewidth=2pt,topline=true,%
frametitleaboveskip=\dimexpr-\ht\strutbox\relax
}
\begin{mdframed}[roundcorner=5pt,nobreak=true]\relax%
\label{#2}}{\end{mdframed}}

\begin{document}

\markboth{S. Knapen and S. Lowette}{Long-Lived Particles at the LHC}

\title{A guide to hunting long-lived particles at the LHC}

\author{Simon Knapen$^{1,2}$, Steven Lowette$^3$
\affil{$^1$Lawrence Berkeley National Laboratory, Berkeley, United States, 94720; email: smknapen@lbl.gov}
\affil{$^2$Berkeley Center for Theoretical Physics, UC Berkeley, Berkeley, United States, 94720}
\affil{$^3$Inter-university Institute for High Energies (IIHE), Vrije Universiteit Brussel, Brussels, Belgium; email: steven.lowette@vub.be}}

\begin{abstract}
This article is a pedagogical review to searches for long-lived particles at the LHC, primarily aimed at experimentalists and theorists seeking to initiate and/or deepen their research in this field. We cover general theory priors and example models, the main experimental techniques for long-lived particles and some of the subtleties involved with both estimating signal efficiencies and background rates.
\end{abstract}

\maketitle

\tableofcontents

\section{Introduction}

The discoveries of the muon (1936) \cite{Neddermeyer:1937md} and the kaon (1947) \cite{Rochester:1947mi} marked the emergence of particle physics as a new field of physics, as they were the first novel particles which did not fit in the contemporary framework of atomic and nuclear physics. Their discoveries famously hinged on the macroscopic lifetimes of both particles, which could be resolved beautifully with the cloud chamber technology of the early 20$^\text{th}$ century. As the old cloud chambers gave way to modern day silicon trackers and time projection chambers, the ability  to accurately measure a decay length has remained an important tool in particle spectroscopy, on par with accurate energy and momentum measurements. In the era of large particle accelerators, the focus shifted towards achieving higher center-of-mass energies and luminosities, while maintaining good energy and momentum resolution. This approach led to the discovery of the top quark and the $Z$, $W$ and Higgs bosons, all of which decay promptly. Measuring particle lifetimes nevertheless remained critical, in particular for $b$-tagging purposes, and a number of important searches for exotic long-lived particles were carried out both at LEP (see e.g.~\cite{DELPHI:1998bkd,Heister:542671}) and the Tevatron (see e.g.~\cite{PhysRevD.78.032015,PhysRevLett.97.161802}).

Given these historical precedents, exotic long-lived particles (LLPs) were always on the radar as a prime discovery mode for beyond the Standard Model physics at the LHC. In the pre- and early-LHC era, this primarily manifested itself in the form of signals of supersymmetry, the dominant theoretical paradigm at the time. It is indeed remarkable how wide of a range of long-lived signatures even ``vanilla'' supersymmetry can generate. One feature supersymmetric signals have in common however, is that they tend to be relatively \emph{hard}, in the sense that they come with one or more high energy photons, jets or leptons. While there certainly existed models early on which generated much softer displaced signatures \cite{Strassler:2006im}, this possibility was not yet as mainstream as it is today, both for practical reasons and because of the aforementioned theory priors before Run-1 of the LHC.\footnote{ As of today, the LHC's data taking schedule is divided in runs with in between long shutdowns for maintenance and upgrades: Run-1 (2010-2012, 7 and 8 TeV), Run-2 (2015-2018, 13 TeV) and Run-3 (2022-2025, 13.6 TeV). From 2029 onwards, the high-luminosity phase of the LHC project will start, the HL-LHC.} The second major evolution within the theory community has been an increased awareness of experimental subtleties, as recasting existing LLP searches or proposing new ones required a much more detailed understanding of the capabilities and limitations of modern trigger and reconstruction algorithms, as well as an intuition for the often very subtle backgrounds associated with the searches.

Since the start of the LHC, the ATLAS, CMS and LHCb collaborations have risen to the challenge and have produced an impressive set of new LLP results, often relying on highly innovative strategies. This was made possible because the lessons of the past have not been forgotten, despite the fact that the primary focus during the design phase of the ATLAS and CMS detectors was on the Higgs and SUSY discovery potential. Due to their hermeticity and excellent tracking capabilities, both experiments have proven to be powerful multipurpose detectors for long-lived signatures. They are further complemented by the superb tracking and vertexing capabilities of the LHCb detector in the forward regime. In those instances where the detector design appeared suboptimal for the signature of interest, the ingenuity of the analysis teams has enabled the collaborations to greatly surpass their design sensitivity. In some cases this was done by extending techniques originally developed as probes of the SM, such as $b$-tagging. Today, the LLP program is a major component of the LHC program as a whole, and experimental techniques that were once niche, such as $dE/dx$, displaced tracking or time-delayed signals, are now powerful and common tools in the experimentalist's arsenal.

In this review we aim to equip graduate students and postdocs entering the fascinating field of LLPs with the tools of the trade, both on the theory and experimental side. By reading our review, we hope young theorists will learn the many new variables and experimental techniques that are needed in searches for long-lived particles, as well as develop a good intuition for what is and is not reasonable experimentally in terms of triggers, reconstruction and backgrounds. Experimentalists getting started on LLPs will hopefully gain a better understanding of the modern theory priors, which types of models are currently on the market and which features they can and cannot predict. Finally, we hope that this text may also serve as a convenient bridge point for more senior colleagues on both sides of the LLP effort, by summarizing the main physics points within a common and mutually intelligible vocabulary.

\section{Theory perspectives}

In this section, we briefly review some general theory priors and lessons from the Standard Model (SM), followed by some explicit examples. A more comprehensive overview of models featuring LLPs can be found in \cite{Lee:2018pag} and \cite{Curtin:2018mvb}.

\subsection{General theory priors\label{sec:modelgeneral}}

As argued in the introduction, our (recent) historical focus on prompt signatures has an excellent theoretical and phenomenological justification: In all known examples, the natural width of a particle with mass $m$ can be estimated by $\Gamma \sim m/8\pi $, unless one or more of the following is true:
\begin{enumerate}
\item The decay occurs through a heavy, off-shell particle, which implies that the width of the LLP is suppressed by a factor of $(m/M)^{2\#}$ with $M$ the mass of the heavy, off-shell state and $\#$ a positive integer that depends on the symmetries of the theory. In the SM, the role of the heavy scale $M$ is usually played by the $W$ mass ($m_W$). The $\pi^{\pm}$, muon and kaons are clear examples.
\item The decay is subject to a severe phase space suppression, because the sum of the masses of the final states is very close to the mass of the LLP. The most spectacular example in the SM is that of the neutron. 
\item The decay width is suppressed by a very small dimensionless coupling constant, associated with a high quality, approximate symmetry. While the smallness of $V_{cb}$ in particular plays a role in some decays, there is no example in the SM for which a macroscopic lifetime can be understood solely in terms of a small coupling constant or mixing angle. 
\end{enumerate}

\begin{figure}
\includegraphics[width=\textwidth]{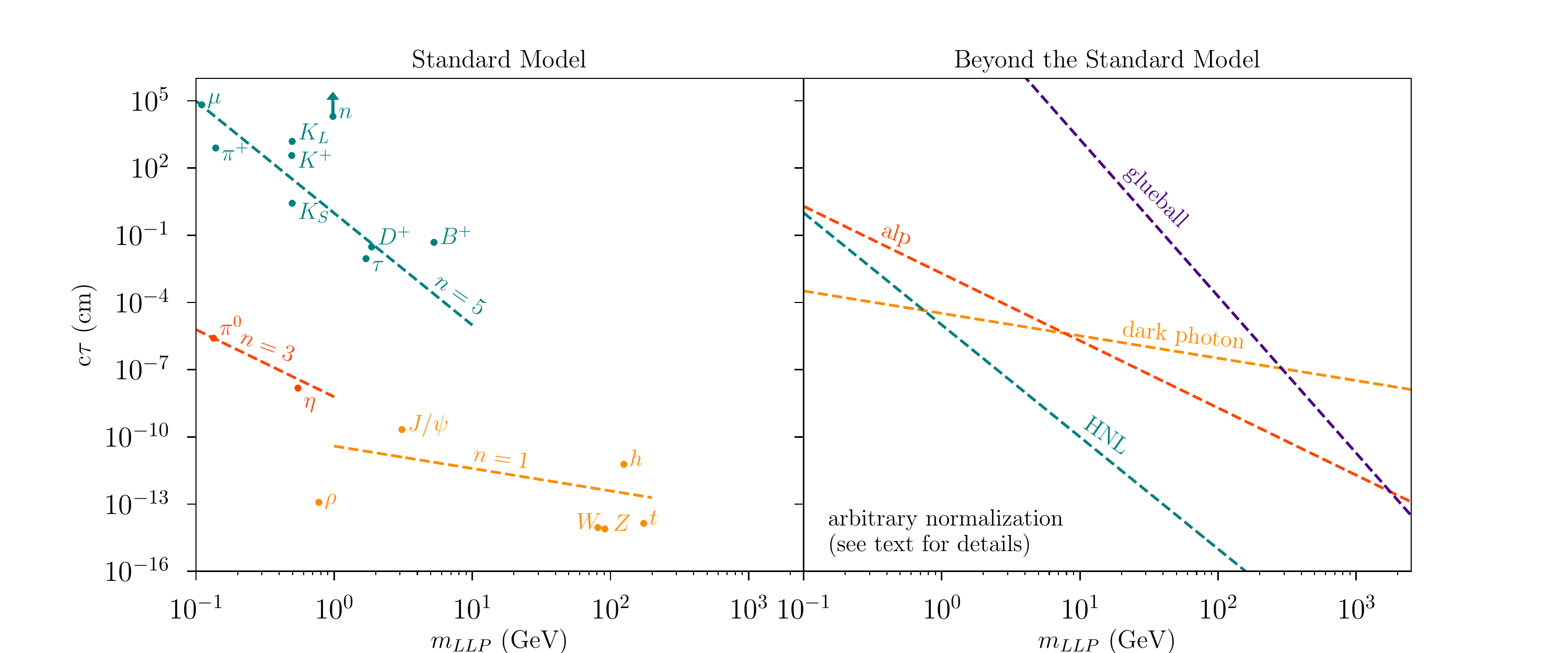}
\caption{\textbf{Left:} Proper lifetime versus mass for selected SM particles, grouped according to the index $n$ in equation \eqref{eq:schematic}. The neutron is the clear outlier due to its extreme phase space suppression; to keep the axis range manageable, the neutron point was brought down and labeled with an arrow to indicate that its true $c\tau$ is much larger than depicted here. It was also excluded from the fit that resulted in the blue curve. \textbf{Right:} Four examples of BSM models, where the dashed lines indicate the approximate scaling with their mass. (Small steps due to kinematic thresholds were neglected.) The overall normalization for each line was chosen to roughly correspond to the \emph{smallest} $c\tau$ allowed by current constraints, but is otherwise arbitrary. The figure is therefore \emph{not} meant to illustrate that large $c\tau$ at high $m_{LLP}$ are disfavored. 
\label{fig:LLPoverview}}
\end{figure}

We can summarize these points in the schematic formula
\begin{equation}\label{eq:schematic}
\Gamma \sim \frac{ \epsilon^2}{(8\pi)^{a-1}} \frac{m^n}{M^{n-1}}
\end{equation}
where $n$ is always an odd, positive integer. The parameter $a$ is also a positive integer and indicates the number of final state particles, while $\epsilon$ represents a potentially small, dimensionless parameter, such as a Yukawa coupling, CKM matrix element or mixing angle. The toy formula clearly reflects the three suppression factors discussed above, if we interpret $m$ as the scale characterizing the volume of the phase space. Usually, this means identifying $m$ with the mass of LLP. For e.g.~the neutron decay one would instead identify $m\equiv m_n-m_p-m_e\ll m_W$, which explains its extra-ordinary lifetime. Equation \eqref{eq:schematic} therefore allows us to group the particles in the SM into \emph{equivalence classes}, indexed by the integer parameter $n$. The higher $n$, the more long-lived the particle and the steeper the mass-lifetime dependence. This is illustrated in the left-hand panel of Fig.~\ref{fig:LLPoverview} for a handful of example particles in the SM. The dashed trend lines are fits to the data, where we held $n$ fixed but floated the intercept.

For the $\rho$, $J/\psi$, top, Higgs, $W$ and $Z$, none of the above suppression factors apply and they naturally decay promptly. The $\pi^0$ and $\eta$ decay to photons through the chiral anomaly, which induces the scaling $\Gamma \sim m^3/f^2$ with $f$ the scale of chiral symmetry breaking. Finally, the $\mu$, $\tau$ and flavored mesons all decay through an off-shell $W$, hence the $\Gamma \sim m^5/m^4_{W}$ trend. The neutron is the clear outlier due to its enormous phase space suppression.

Moving on to extensions of the SM, equation \eqref{eq:schematic} still applies. We can thus deploy it to critically examine a few pieces of ``conventional wisdom'' which are sometimes applied to LLP models. When inspecting the left-hand panel of Fig.~\ref{fig:LLPoverview} without accounting for the color coding, one may be tempted to conclude that lifetime simply correlates with mass. This is the origin of a commonly held intuition:
\begin{loreno}[Misleading]{lore:special}
Heavy particles prefer to decay promptly, while light particles can have long lifetimes. Heavy LLPs therefore require ``special'' models and/or parameter choices.
\end{loreno}
While the proper lifetime ($c\tau$) undoubtable scales inversely with the particle mass, it is immediately obvious from equation \eqref{eq:schematic} that the above statement is too hasty: Since $n=5$ is pretty typical in the SM and beyond, even a moderate hierarchy of scales in our BSM model suffices to suppress the decay width by many orders of magnitude. The most famous example is presumably the long-lived gluino in a split-SUSY scenario \cite{Arkani-Hamed:2004ymt}, which arises from the $m_{\tilde g}\ll m_{\tilde q}$ hierarchy.\footnote{As is the case in the SM, new LLP's may be charged or neutral, or even colored in the case of long-lived gluinos. Charged and/or colored LLPs are prevalent in supersymmetry in particular, while neutral LLPs tend so show up in models of dark matter, baryogenesis and certain non-SUSY solutions the hierarchy problem. 
} Heavy neutral leptons (HNLs) \cite{Abdullahi:2022jlv} with mass $m\ll m_W$ are another example in this same equivalence class, as shown in the right-hand panel of Fig.~\ref{fig:LLPoverview}. In fact, just as for the SM, \emph{all} BSM LLPs can be classified according to the index $n$. In Fig.~\ref{fig:LLPoverview} we show the dark photon ($n=1$), the axion-like particle (ALP, $n=3$), HNL ($n=5$) and dark sector $0^{++}$ glueballs ($n=7$) \cite{Craig:2015pha}.

This brings us to the second piece of common wisdom:
\begin{loreyes}[Largely true]{lore:theorists}
Theorists are not so good at predicting $c\tau$.
\end{loreyes}
While there are important exceptions in very simple models such as HNLs or dark photons, this is largely true, but not because of lack of perseverance or cleverness: The heavy mass scale $M$ can easily be outside the reach of the LHC and can be difficult to pin down theoretically. In some models it can be estimated by using additional inputs, such as the dark matter relic density, but even in those cases, even modest uncertainties on $M$ can get amplified greatly in equation \eqref{eq:schematic}. This gets progressively worse for larger $n$, with some composite particles such as dark sector glueballs as the most extreme examples. Fortunately, we will see in Sec.~\ref{sec:geoacceptance} that our inability to predict $c\tau$ more precisely is actually not so relevant experimentally.

All that said, there is a universal and relatively model-independent upper bound on $c\tau$ from potentially spoiling Big Bang Nucleosynthesis (BBN) \cite{Fradette:2017sdd}. The argument goes as follows: Any particle with an observable direct production at the LHC should at some point in time be in thermal equilibrium with the cosmic plasma, as long as the universe was hot enough to produce it. As the universe cools, the LLP freezes out from the SM thermal bath and subsequently decays. If this decay occurs while BBN is happening, the injection of additional particles and energy tends to modify the primordial abundances, which tends to be in conflict with observation. This gives us a range of upper bounds between \mbox{$\tau\lesssim 0.1$ - $10^4$ s}, depending on the dominant decay modes and the abundance at the onset of BBN. While interesting for its universality, even 0.1 s is an absurdly long time scale for any collider experiment, and this bound is therefore almost never experimentally relevant. Lower bounds on $c\tau$ also exist, but they are always dependent on the model. This does \emph{not} mean however they are always trivial to evade, and they must be investigated on a case-by-case basis. Following equation \eqref{eq:schematic} and Fig.~\ref{fig:LLPoverview}, such lower bounds tend to exist for low $m$ ($m\lesssim 5$ GeV) and $n\geq 5$. Concretely, a smaller $c\tau$ would require a larger $(m/M)^n$, while model-dependent lower bounds on the scale $M$ exist from direct searches for other particles in the model. Important examples are HNLs and any light LLP that is a composite particle, e.g.~in hidden valley models~\cite{Strassler:2006im,Knapen:2021eip}.

\subsection{Example models}

Rather than supplying a comprehensive overview of models, which exists elsewhere \cite{Lee:2018pag,Curtin:2018mvb}, here we touch upon three example models, one for each of the three general principles laid out in the previous section. We deliberately picked examples which may be a little bit less well-known than the classic examples (e.g.~split SUSY), which are discussed abundantly elsewhere. 

The most minimal way of obtaining a macroscopic lifetime is to have the SM $W$-boson serve as the heavy, off-shell particle. This is precisely what happens with \textbf{heavy neutral leptons}, which mix with the SM neutrinos and undergo a 3-body decay to a SM neutrino plus two other SM fermions. It is therefore no surprise that its lifetime obeys the $c\tau \sim m^{-5}$ scaling law in Fig.~\ref{fig:LLPoverview}, similar to the SM flavorful mesons and heavy leptons. It also explains why with similar values for their mixing angles and masses, the lifetime of a low mass HNL is much longer than that of a dark photon. For a recent and comprehensive overview of HNL phenomenology we refer to \cite{Abdullahi:2022jlv}.

A strong phase space suppression can be the result of an approximate symmetry \cite{Sher:1995tc}, as is the case in the  long-lived chargino models that generate the famous disappearing track signature. It can also be motivated from cosmological considerations however: Suppose we have two dark sector particles $\chi_1$ and $\chi_2$ with $m_{\chi_1}<m_{\chi_2}$, which can annihilate to SM fermions $(f)$. Since dark matter ($\chi_1$) freeze out tends to happen at a temperature $T\lesssim m_{\chi_1}/10$, the $\chi_1 \chi_1 \to f\bar f$ freeze-out process receives an $\sim \exp(-2m_{\chi_1}/T)$ Boltzmann suppression in the early universe. However, if a \textbf{co-annihilation} process $\chi_1 f \to \chi_2 f$ exists, it is only suppressed by $\sim \exp(-m_{\chi_2}/T)$ and could therefore easily dominate if $m_{\chi_2}-m_{\chi_1}\ll m_{\chi_1}$  \cite{PhysRevD.43.3191}. The presence of a second particle $\chi_2$ that is close in mass to $\chi_1$ can therefore radically change the relic abundance of $\chi_1$. Moreover, if $\chi_2$ can be produced at a particle collider, its decay ($\chi_2\to \chi_1 f\bar f$) can be very slow, due to the small splitting between $\chi_1$ and $\chi_2$ \cite{Bai:2011jg,Izaguirre:2015zva}.

Though this possibility is not realized in the SM, there are many BSM models in which a particle acquires a macroscopic lifetime because of a very small coupling. \textbf{WIMP baryogenesis} is such a model \cite{Cui:2012jh,Cui:2014twa}. The idea is that a WIMP ($\chi$) freezes out in the early universe, but instead of being the dark matter, it is allowed to decay to SM baryons and the actual dark matter ($X$). If baryon number, C and CP are all violated in the decay, we expect $\text{Br}(\chi \to B X)\neq \text{Br}(\chi \to \bar B \bar X)$, where $B$ ($\bar B$) represent a SM final state carrying (anti)-baryon number. If the decay is fast as compared to the expansion rate of the universe, the inverse process where the $B$ and $X$ fuse back into $\chi$ is also efficient, and we end up with no net asymmetry. We must therefore assume that $\chi$ is a long-lived particle in the context of the early universe. It turns out that for $m_\chi$ in the hundreds of GeV range, this implies that $\chi$ is also long-lived at the LHC. 

\section{Experimental signatures}

The experimental signatures we can use  to distinguish LLPs from backgrounds strongly depends on the properties of the LLP: the $c\tau$, the decay modes, the LLP mass, charge, etc. At the same time, the experimental capabilities of the detectors provide the boundary conditions under which these signatures can be used for signal selection. Luckily, the LHC detectors were instrumented for heavy flavor identification, have accurate timing capabilities, and possess other features useful for LLP detection - they are thus very well suited to search for LLPs.
\begin{loreno}[False]{lore:detector}
The LHC detectors were not equipped for LLP detection.
\end{loreno}
This interplay of the characteristics of the LLPs with the experimental context drives the selection choices made in data analysis. In the following, we overview the various experimental tools at the disposal of the intrepid explorer hunting for new LLPs.

\subsection{Displaced tracks and vertices}

Charged particles from the decay of an LLP are excellent experimental probes  that can be used to discriminate a potential long-lived new-physics signal from the SM debris of the collisions, for which the charged particles mostly emerge promptly from the collision point.
One of the experimental limitations restricting the identification of LLPs is the capability for the reconstruction of displaced charged particles. This reconstruction of displaced tracks is a trade-off challenge between efficiency and purity, with computing capabilities restricting the ultimate performance. We discuss displaced track reconstruction in more detail in Section~\ref{sec:offlinereco}.
Starting from the displaced tracks, there are several handles at our disposal for LLP identification.

\begin{figure} 
\begin{minipage}{.5\textwidth}
 \includegraphics[height=4cm]{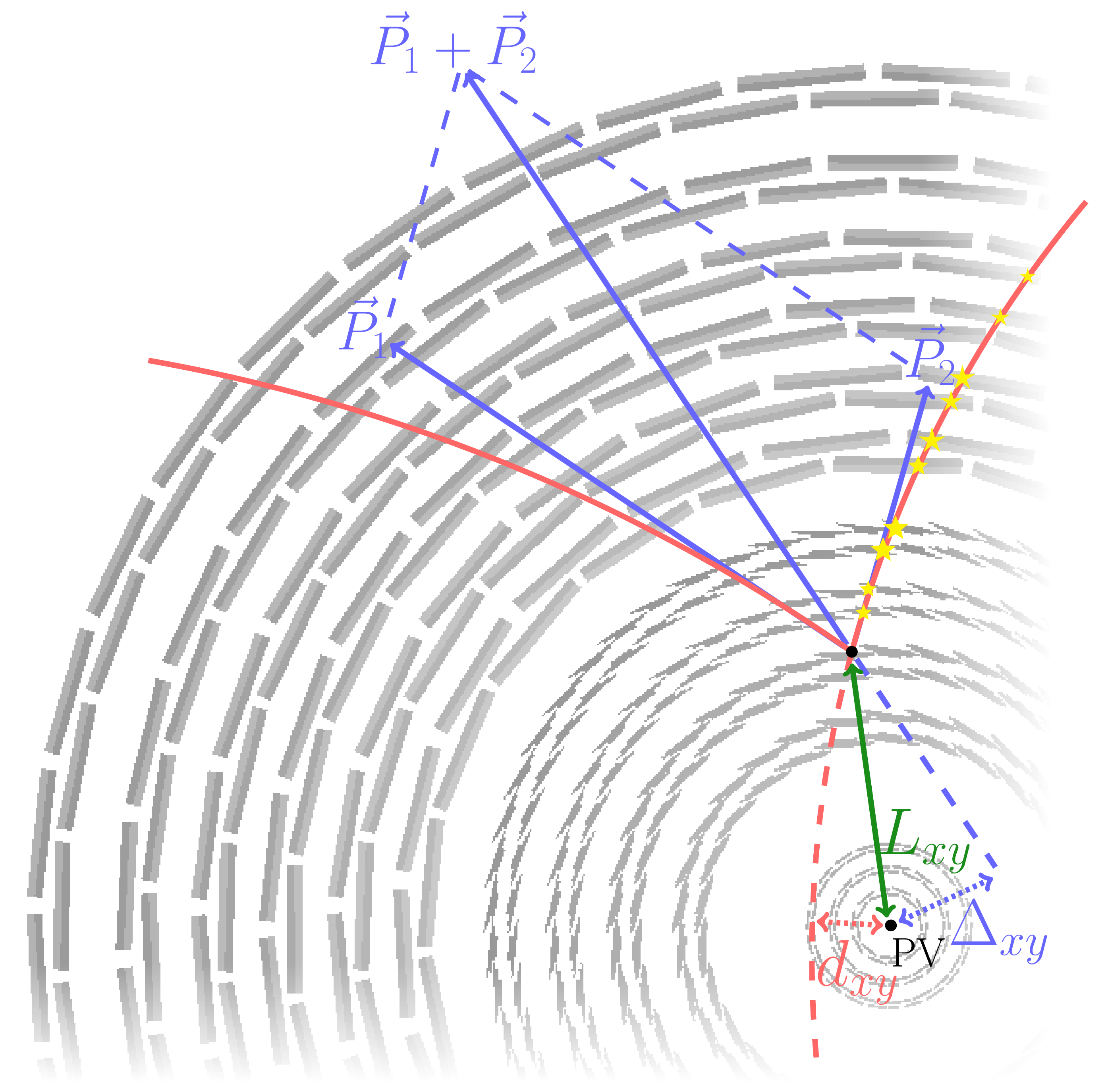}
 \end{minipage}%
\begin{minipage}{.5\textwidth}
 \includegraphics[height=4cm]{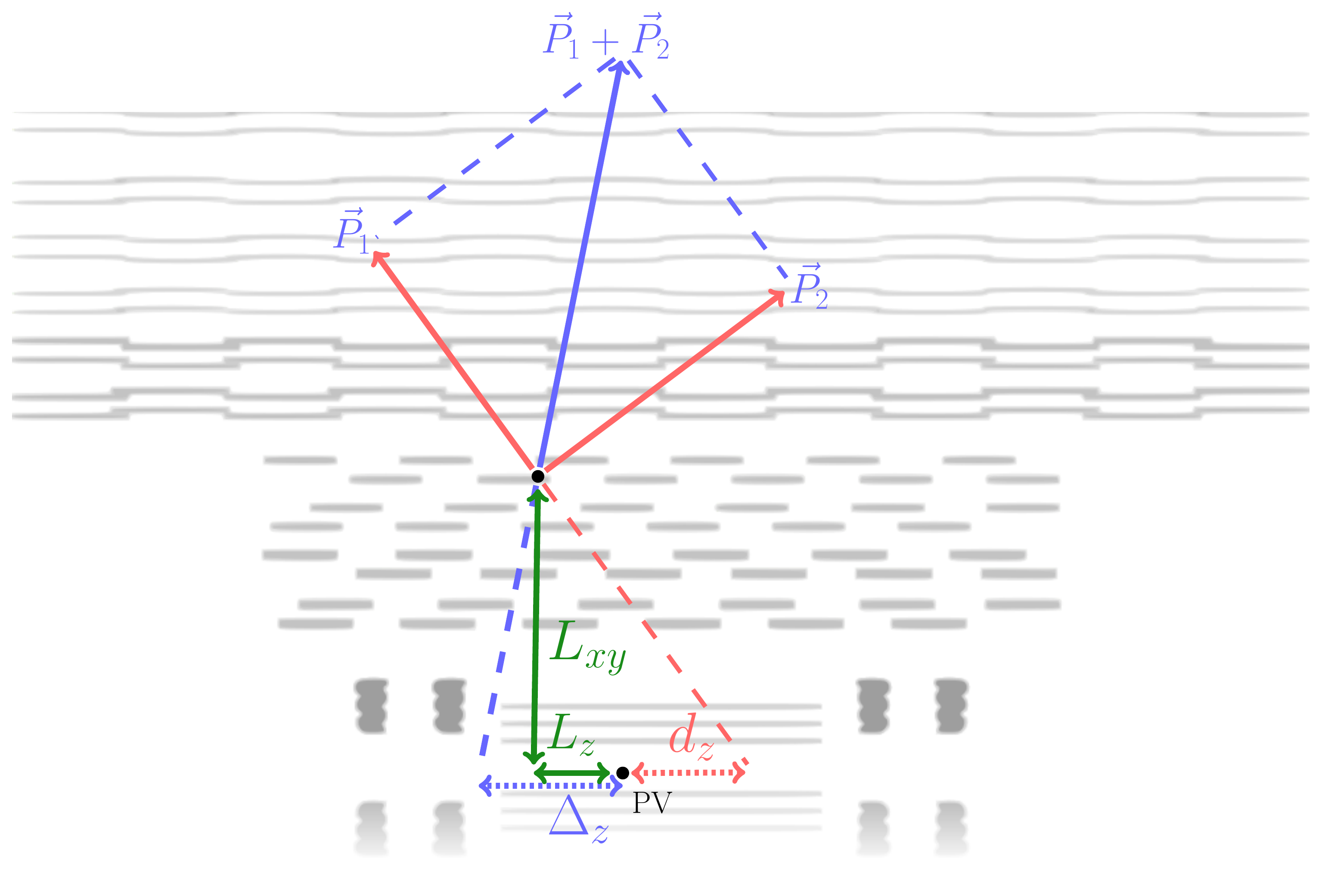}
\end{minipage}
\caption{Drawing of the transverse (left) and longitudinal (right) cross sections of the current CMS inner tracker, with an example displaced vertex and the variables defined in the text. The energy loss ($dE/dx$) in individual layers can fluctuate significantly, as indicated schematically by the varying sizes of the yellow stars.\label{fig:trackvars}}
\end{figure}

\begin{description}

\item[Track impact parameters]

The impact parameters of a track are defined by the vector from the collision point, called the primary vertex (PV), to the closest approach of the track to that point. Often this is split in its transverse component, the length of which is referred to as $d_{xy}$, and similarly $d_z$ for the longitudinal component. In Fig.~\ref{fig:trackvars}, a graphical representation is given of $d_{xy}$ and $d_z$ for a displaced track, depicting the extrapolation towards the primary vertex.

The impact parameter is a simple yet powerful discriminator between prompt and displaced tracks. Before the advent of more powerful multivariate techniques, $d_{xy}$ was a main ingredient in the identification of jets from heavy flavor, targeting the identification of displaced tracks coming from $B$ and $D$ meson decays. For this reason, the inner tracking detectors of modern particle physics detectors are designed for excellent $b$-jet identification, with a spatial segmentation in the inner silicon pixel detectors as small as $50 \, \mu m$.

The resolution of the $d_{xy}$ and $d_z$ observables is driven by the detector granularity of the detector layers the charged particle crosses first, but also by the distance of the extrapolation to the beamline and by the number of hits on the track. Thus, the $d_{xy}$ resolution of a track that emerges far from the collision vertex may be rather poor. This is in particular the case for tracks reconstructed in the muon spectrometer only. To account for this track-dependent varying resolution, some analyses employ the significance of the impact parameter, $d_{xy}/\sigma_{xy}$ and $d_z/\sigma_z$, where $\sigma_{xy}$ and $\sigma_z$ are estimates of the impact parameter uncertainties obtained from the track fit. These significance quantities are more performing, at the expense of requiring more care when modeling or reproducing their efficiencies and related systematic uncertainties.

For prompt background tracks, the impact parameter resolution $\sigma_{xy}$ can reach below $20\,\mathrm{\mu m}$ for tracks with transverse momenta above $10\,\mathrm{GeV}$, but is more of the order of $100\,\mathrm{\mu m}$ for tracks with $p_{\rm T} \sim 1\,\mathrm{GeV}$ or large $\eta$~\cite{CMS:2014pgm,ATLAS:2015ngk}. In the $z$ direction, where the pixels are elongated, the resolution is somewhat worse, reaching at best $\sigma_z\sim30\,\mathrm{\mu m}$ for high-momentum tracks. Another complication along the $z$-direction is the spread of the collision vertices with a standard deviation of about $11\,\mathrm{cm}$, while the transverse beam size is at the level of $25\,\mathrm{\mu m}$. An accurate determination of $d_z$ thus relies on an unambiguous identification of the primary vertex in the presence of many simultaneous proton collisions, known as \emph{pileup}. Often analyses will not make use of $d_z$ for these reasons, but only use the already very powerful displaced track identification handles $d_{xy}$ and $d_{xy}/\sigma_{xy}$.

\item[Track and vertex displacement]

If the LLP is light and has a high momentum, the tracks may emerge well displaced but still have a small impact parameter and low significance. At that point, the transverse ($L_{xy}$) and longitudinal ($L_z$) displacements of the decay vertex, shown in Fig.~\ref{fig:trackvars}, may be better discriminators. For a displaced, isolated single track, the starting point cannot be unambiguously determined, but the missing inner hits still hold valuable information. In the case of two or more tracks emerging from the same LLP decay, a vertex can be fit from the tracks, yielding an often accurate 3-dimensional estimate of the decay point, and thus of $L_{xy}$ and $L_z$. Also here the dimensionless significance is sometimes preferred. In particular in the case of a decay with collimated displaced tracks, the position of the vertex in the direction of flight may be difficult to estimate, and the normalization of the displacement to its uncertainty can provide improved discrimination against poorly measured backgrounds.

\item[Track multiplicity and vertex mass]

Another discriminator that may be of use to select particular signals such as displaced hadron jets is the track multiplicity at a displaced vertex. The inefficiency of displaced tracking reconstruction, increasing with the displacement of the decay vertex, makes this variable sensitive to mismodeling in simulation. As a result, a requirement of a minimum or maximum number of tracks can induce a significant signal selection uncertainty.
Nevertheless, a minimum number of tracks is often imposed as a preselection (having a vertex already implies at least two) as it strongly suppresses some of the backgrounds (see Sec.~\ref{sec:background}). Alternatively, also vertex mass (i.e. the invariant mass of the tracks that form a vertex) is a related observable that can be used for this purpose.
Examples of the use of both variables can be found in Refs.~\cite{ATLAS:2017tny,ATLAS:2022fag,CMS:2020iwv,LHCb:2021dyu}.

\item[Decay direction]

For new-physics scenarios where the LLP decays to charged as well as neutral particles, e.g. in the case of displaced tau leptons, the jet clustered from the charged particles may point off-axis with respect to the LLP flight direction. In such cases, both the impact parameters ($\Delta_{xy}$ and $\Delta_z$ in Fig.~\ref{fig:trackvars}) and the displacement of the jet can be used for signal selection, and in case of discovery the direction information from the displaced tracks can be used to also characterize the neutral component. The usefulness of this observable is expected to be strongly decreasing for larger displacements, as tracking inefficiencies will wash out possible sensitivity from the decay direction. For signals without neutral particles in the final state, $\Delta_{xy}$ and $\Delta_z$ are expected to be zero, within the resolution. For those signals they can therefore be useful variables to suppress backgrounds from fake vertices (see Section~\ref{sec:fakes}), as those tend to produce a flat distribution in $\Delta_{xy}$ and $\Delta_z$ (see e.g.~\cite{CMS:2021sch}).

\item[Ionization loss] 

Tracking detectors such as time projection chambers or silicon trackers can also measure ionization energy deposits per unit length along the track ($dE/dx$).\footnote{Though the units of $dE/dx$ are MeV/cm, it is usually reported in units of MeV$\times \text{cm}^2/g$ in the Particle Data Book \cite{ParticleDataGroup:2022pth}. In these cases one has divided out by the mass density of the target, as this allows for a more consistent definition for gas targets.} For new charged particles with masses well beyond the charged pion, kaon or proton mass, or with electric charge different from $1e$, the ionization deposits provide an additional handle to discriminate such signals from regular tracks from SMbackgrounds.

The strongest discrimination may be achieved for particles with a boost ($\beta\gamma$) that places them below the usual minimum ionizing plateau of the Bethe-Bloch curve. It is possible to discriminate  different SM hadron or ion species in this manner, but only at very low momenta, of the order of $1 \, \text{GeV}$ (see e.g.~supplementary material of~\cite{ATLAS:2018lob}).
New heavy stable charged particles (HSCP's), such as R-hadrons, can have masses in the multi-TeV regime and have therefore a low enough boost to allow discrimination with $dE/dx$ measurements \cite{ATLAS:2022pib,CMS:2016kce}. For new particles with non-unit electric charge, on the other hand, the ionization mean free path dependence on the square of the charge makes the ionization loss a strong discriminator for a much broader mass range \cite{ATLAS:2018imb}. Searches for ultra-highly ionizing particles such as magnetic monopoles are the most extreme examples in this category \cite{ATLAS:2019wkg}.

While ionization of material by particles is a very well understood phenomenon~\cite{ParticleDataGroup:2022pth}, the accurate simulation of the background particles that are produced in the collisions, as well as the description of the detector geometry, material, aging with radiation, electronics saturation, etc, render a precise simulation of ionization loss in the detector very difficult for backgrounds as well as signals. Analyses using ionization loss as a discriminator between signal and background thus need to carefully calibrate energy loss using e.g.\ hadron tracks or muons from $Z$ bosons, often combined with background predictions extracted directly using data. 

Simulations of $dE/dx$, whenever possible, are performed with very sophisticated software packages such as GEANT4 \cite{ALLISON2016186} and may therefore not be practical for theorists seeking to perform a reinterpretation. It is therefore tempting to just rely on the Bethe-Bloch curve, which provides the \emph{mean} energy loss of a particle through matter. This leads to an important pitfall however, as the energy loss is a stochastic process with a highly skewed distribution, such that the mean energy loss is dominated by rare, high energy collisions. The \emph{most probable} energy loss is often a more useful estimator~\cite{ParticleDataGroup:2022pth}, especially for thin detectors such as the tracking layers.\footnote{Since energy loss in materials is a very subtle matter,  we advise  theorists seeking to use analytic formulas to model $dE/dx$ to verify all the limitations of the formulas provided in the relevant chapter in Ref.~\cite{ParticleDataGroup:2022pth}.}

\item[Track timing]

Precise time measurements of the energy deposits in the detectors provide another source of information on charged particles. Inner tracking detectors, close to the beamline, do not provide a direct estimate of the peak time of the hits, and are typically  read out in a narrow time window around the expected arrival time for particles traveling at or close to the speed of light \mbox{($\beta = v/c = 1$)}. For moderate $\beta$, this implies that an apparent smaller ionization is recorded, though estimating a time delay from these stochastic measurements is typically imprecise and further confounded by the unknown LLP mass and charge.
Track timing is more relevant for particles traversing the outer muon chambers, where the individual cluster timing measurements are controlled to the level of about $2\,\mathrm{ns}$~\cite{Chiodini:2013rta,CMS:2013vyz}. 
Furthermore, the track fit can be improved by having the hit position estimates from the gas-ionization signals to be dependent on $\beta$. The accurate measurement of the particle's speed is leveraged by the long travel distance in the detector, which makes it possible to reach a resolution on $\beta$ as small as 5\%~\cite{ATLAS:2011zgn,CMS:2016kce}. 

\end{description}

\subsection{Calorimeter signals}
The reconstruction of calorimetric signatures of LLPs is not hampered by a low efficiency, as can be the case for the displaced tracking, but the lack of tracks means that identifying LLP-induced energy deposits is very non-trivial. There are however a number of powerful, advanced experimental handles, as described in the following.

\begin{description}

\item[Delayed calorimeter signals]

Calorimeters measure particle energies by observing scintillation light arising from electromagnetic or hadronic showers induced by the interactions of the  incoming particle with an absorber.
Depending on the calorimeter design, accurate measurements of the signal timing are available. In ATLAS and CMS, the electromagnetic calorimeters (ECAL) have an ideal, intrinsic time resolution as low as about $70\,\mathrm{ps}$ for energies larger than several tens of $\mathrm{GeV}$~\cite{delRe:2015hla,Mahon:2020hyp}. In practise, the time resolution for measurements of energetic photons in the barrel is at best of the order of $200\,\mathrm{ps}$~\cite{Mahon:2020hyp,CMS:2013lxn}, dominated by a component arising from the longitudinal spread of the LHC beams.
For the hadron calorimeters (HCAL) the timing resolution is also rather accurate. The ATLAS Tile Calorimeter achieves a resolution as low as $0.4\,\mathrm{ns}$ for high energy deposits~\cite{ATLAS:2018edp}, while CMS reports a time resolution of its HCAL of $1.2\,\mathrm{ns}$ for jet energies above $100\,\mathrm{GeV}$~\cite{CMS:2009nwd}. 

These rather precise time resolutions make it possible to use also the calorimeters to distinguish background low-mass ultrarelativistic particles from particles with low $\beta$ or a delayed signal from an increased path length from a displaced decay. While a low $\beta$ easily induces multi-$\mathrm{ns}$ delays for new particles with large masses, e.g.~in the $\mathrm{TeV}$ range, also the path length can bring several $\mathrm{ns}$ delay, well above the timing resolution~\cite{CMS:2019qjk}.

\item[Displaced calorimeter signals]

The identification of signals unusually directed or displaced inside a calorimeter is another powerful experimental handle on LLPs decaying in the detector. Although calorimeters are often sampling the developing shower in several alternating layers of absorber and scintillator, a detailed segmentation of the shower is not always available for offline analysis. This is particularly the case for the CMS HCAL, because the detector design aggregates measurements along the shower depth. For most purposes, the shower energy estimate suffices, but for LLP identification \mbox{3-dimensional} information on the shower profile is desired. It can then be used to search for decays happening deep inside the calorimeter using e.g.~the absence of an ECAL energy component in a hadronic shower, or the detailed calorimeter cluster depth position~\cite{ATLAS:2016krp,ISIK2022167389}. Furthermore, fine-grained calorimeter shower information may help identify photons that hit the calorimeter under an angle due to a displaced decay, using the photon's direction measurement~\cite{ATLAS:2022vhr} or its elliptical shower shape~\cite{CMS:2019zxa}.

\end{description}

\subsection{Other experimental handles\label{sec:otherexphandles}}

Beyond the basic functionalities described above, detectors also start being used well beyond their initial design. The recent CMS search  for hadronic decays in the muon detectors is an excellent example \cite{CMS:2021juv}: Thanks to the thick steel of the magnet's return yoke, the CMS muon system can be used as a calorimeter, clustering low-level muon chamber hits in the muon detectors. A related search also exists in ATLAS \cite{ATLAS:2022gbw}, which has a larger fiducial volume in its muon detector, but has air as opposed to steel in between its tracking detectors. As a result, ATLAS relies on reconstructing displaced tracks and a displaced vertex, as opposed to searching for a calorimetric shower. This means that the efficiency of the ATLAS search is more sensitive to the mass of the LLP, while the CMS search primarily  depends on its energy. 
Combining information across subdetectors can also shed light on the individual measurements, as recently exemplified in Ref.~\cite{ATLAS:2022pib}.

More detailed particle identification can furthermore be a powerful tool in specific exclusive searches, but is harder to achieve without dedicated detectors. An example here is the use of the LHCb RICH detector to identify slow moving HSCPs \cite{LHCb:2015ujr} and charged kaons coming from displaced low-mass scalar decays~\cite{CidVidal:2019urm}. With the advent of the HL-LHC, we will see an expansion of the experimenter's portfolio with additional sub-detectors and detector capabilities. Most prominently, dedicated precise timing detectors with resolutions down to a few tens of ps are being added in front of the calorimeters~\cite{CMS:2667167,CERN-LHCC-2020-007}. But also many of the other subdetectors are getting upgrades that provide opportunities for inventive new approaches to establish LLP particle signatures (see e.g.~\cite{CERN-LHCC-2017-020,Collaboration:2759072}).

\section{Signal selection}

Quantifying the signal selection efficiency is more complicated for LLPs than for prompt particles, and can be very difficult to simulate and parametrize. This leads to a number of important pitfalls for theorists in particular when trying to model an existing or proposed analysis. For experimentalists it is moreover important to simulate the desired signals as efficiently as possible, due to the high computational complexity of the full detector simulation. In this section we summarize a few tricks, and point out possible subtleties.

\subsection{Signal reweighting and geometric acceptance\label{sec:geoacceptance}}

For prompt particles, the geometric acceptance is typically estimated by requiring that the final states all satisfy a set of relatively simple $\eta$ and $p_T$ cuts associated with the detector geometry. For unstable LLPs this is more complicated, as the trigger and reconstruction efficiencies depend strongly on the location of the decay vertex. Crucially, these efficiencies however do \emph{not} depend on the LLP's proper lifetime ($c\tau$). We can use this to our advantage in a simple reweighting algorithm, as described below.

For simplicity, let us assume that we have a single LLP with momentum $\vec P$ and denote the momenta of its decay products by the set of momenta $\vec p_i$. For a specific model, a sample of  $(\vec P,\vec p_i)$ can be generated with standard Monte Carlo simulation codes, such as MadGraph5\_aMC@NLO \cite{Alwall:2014hca} or PYTHIA 8 \cite{Sjostrand:2014zea}. To calculate the efficiencies for low $c\tau$ correctly, it is very important to model the tail of the LLP's $p_T$-distribution carefully and with enough statistics. Depending on the signal, this can mean including hard initial state radiation (ISR) or simulating events which are weighted based on the LLP's $p_T$. The latter is particularly important for light LLPs which are produced non-resonantly, e.g.~in Drell-Yan or in exotic $B$-meson decays (see e.g.~\cite{Evans:2020aqs}). The reason is that the most boosted LLPs will dominate the efficiency for small $c\tau$.

When generating LLP events with a simulation tool, it is always most efficient to ignore the displacements of LLP decay vertices provided by the event generator, and instead manually generate them along the halfline defined by $\vec P$. In other words, we draw a positive real number $a$ from a probability distribution $f(a)$ and relate it to the distance the LLP traveled before decaying. The choice of $f(a)$ is a priori arbitrary, but should be such that the resulting vertex locations efficiently sample the detector volume of interest. For example, suppose that our sensitive detector element is a co-axial cylinder, e.g.~a calorimeter, and we care only about LLPs for which $L_{xy}^-<L_{xy}<L_{xy}^+$ with $L_{xy}^{\pm}$ the inner and outer radii of the detector. It then makes sense to choose e.g.~the uniform probability distribution
\begin{equation}
f(a)=\frac{1}{L_{xy}^+ - L_{xy}^-} \quad \mathrm{with} \quad L_{xy}^-<a<L_{xy}^+
\end{equation}
and define the vertex location as
\begin{equation}
\vec x = (a\cos \phi , a\sin \phi,a\sinh \eta)
\end{equation}
with $\phi$ and $\eta$ the azimuthal angle and pseudorapidity associated with $\vec P$. This ensures that as few events as possible are wasted on decays outside the fiducial volume. These events can then be passed through the detector simulation to calculate the combined trigger and reconstruction efficiency $\epsilon_r(\vec x,\vec p_i)$ depends only on the vertex location $\vec x$ and final state kinematics $\vec p_i$. (More on this in Sec.~\ref{sec:trigger} and Sec.~\ref{sec:offlinereco}.) This calculation is usually very computationally expensive with a full-fledged detector simulation, but we only need to do it once. With $\epsilon_r(\vec x,\vec p_i)$ in hand, we can obtain the efficiency as a function of $c\tau$ by defining the following weight for each event
\begin{equation}\label{eq:weights}
w(\vec P, \vec p_i,\vec x, c\tau)\equiv \frac{e^{-|\vec x|/\beta\gamma c\tau}/\beta\gamma c\tau}{f(a)} \times \epsilon_r(\vec x,\vec p_i)\times w_{p_T}(\vec P)
\end{equation}
where $\beta\gamma$ is the boost factor of the LLP and $w_{p_T}(\vec P)$ the $p_T$-dependent weight provided by the event generator. The total signal efficiency for a particular $c\tau$ is then defined by simply averaging over the weights
\begin{equation}\label{eq:efffull}
\epsilon(c \tau)=\frac{1}{N}\sum_{\text{events}} w(\vec P, \vec p_i,\vec x, c\tau)
\end{equation}
with $N$ the number of events.\footnote{To generalize the algorithm to multiple DV per event, it suffices to define a weight like equation \eqref{eq:weights} for each DV and combine them in equation \eqref{eq:efffull} with the appropriate combinatorics.}  We thus are able to recycle the same set of events for any value of $c\tau$ we want, greatly reducing the computation cost of the simulation.

To gain an intuition for equation \eqref{eq:weights}, let us return to our simplified calorimeter example of the co-axial cylinder, such that $\epsilon_r(\vec x,\vec p_i)=0$ unless $L^-_{xy}<L_{xy}<L_{xy}^+$. For $\beta\gamma c\tau \ll L_{xy}^-$ we then recover the expected exponential suppression of $\epsilon(c \tau)$, regardless of our choice for $f(a)$. The events with the largest $\beta\gamma$ carry an exponentially larger weight, hence the need to generate $p_T$-weighted events.\footnote{This means that for low enough $c\tau$, the efficiency will always be dominated by a handful of events, rendering the simulation unreliable.  A plot of the cumulative distribution of the weights is a good way to verify whether one is in this regime.} For $\beta\gamma c\tau \gg L_{xy}^+>|\vec x|/(1+\sinh^2\eta)$, we can approximate $e^{-|\vec x|/\beta\gamma c\tau}\approx 1$ which implies that the dependence on $c\tau$ factorizes from the efficiency. In other words
\begin{equation}\label{eq:effscaling}
\epsilon(c \tau) \approx \epsilon(c \tau') \times \frac{c\tau'}{c\tau} \quad \text{if} \quad \beta\gamma c\tau',\beta\gamma c\tau \gg L_{xy}^+.
\end{equation}
This is the reason that the limits from \emph{all} searches for a single displaced vertex scale as $1/c\tau$ in the large lifetime limit, \emph{regardless of the detector geometry and analysis details}. Similarly, one can show that in the long lifetime regime the sensitivity for searches for two independent displaced particles should always scale as $1/(c\tau)^2$. It moreover disproves the following common misconception:
\begin{loreno}[False]{lore:fardetector}
If the detector is close to the interaction point (IP), it is exponentially difficult to detect LLPs with large $c\tau$.  Placing the detector further away therefore leads to a higher signal efficiency.
\end{loreno}
Indeed, equation \eqref{eq:effscaling} shows that it is only \emph{linearly} difficult to detect LLPs with long lifetimes. This is why some LLP searches can be sensitive to $c\tau$ values as large as a kilometer  \cite{CMS:2021juv}. The scaling law in \eqref{eq:effscaling} is moreover independent of the detector location, provided that the typical lab frame lifetime ($\beta\gamma c\tau$) is much larger than the distance between the detector and the IP.  For  low $c\tau$ on the other hand, we saw that the efficiency depends exponentially on the distance between the detector and the IP, such that \emph{a detector as close as possible to the IP always covers the largest range in $c\tau$.} Of course, in practice there may be engineering and/or background considerations which can force one to place detectors further out, as discussed in Sec.~\ref{sec:background} and Sec.~\ref{sec:dedicated}.

\subsection{Triggering on long-lived particles\label{sec:trigger}}

The trigger infrastructure at all LHC experiments is both a critical and very complex component of the experiment. It is responsible for reducing the data intake from 40 MHz sampling frequency down to roughly a kHz, all in very limited time with finite read-out and computing resources. As a result, it is in constant development, through system upgrades, improved algorithms and calibrations, as well as shifting physics priorities. All of this especially affects searches for LLPs, as the event reconstruction is much more complex than for prompt particles. It is therefore important to recognize the trigger's capabilities and limitations before starting an analysis or theory study.

The trigger at ATLAS and CMS consists of two stages: The ``Level 1'' (L1 or LVL1) or ``hardware'' trigger and the ``High Level Trigger'' or ``software'' trigger (HLT). The L1 trigger is responsible for reducing the rate from 40 MHz to maximum 100 kHz, using limited detector information. In particular, there is no time to read out and process the inner tracker data at this stage, and the calorimeters and muon systems are read out at reduced granularity. The L1 trigger therefore makes a decision based on relatively simple observables such as missing transverse energy (MET), the scalar sum of all reconstructed transverse energy ($H_T$),
a high-momentum lepton, etc. Events passing the L1 selection are handed over to the HLT, which reduces the rate further from about 100 kHz to the order of 1 kHz. It can perform a fairly faithful reconstruction of the full event, including tracking information. This allows for more sophisticated selection criteria and there are thus many more trigger paths. Nevertheless, the HLT is currently limited to an average latency of roughly \mbox{250 ms}, which means that a very resource-intensive task such as reconstructing displaced tracks cannot be taken for granted. A good example is that of a purely hadronic displaced vertex in the tracker, for which displaced tracks currently cannot be efficiently reconstructed at HLT. When combined with a moderate $H_T$ requirement, it is however possible to trigger on a jet with an anomalously low number of prompt tracks~\cite{CMS:2020iwv}. If such a prompt-veto strategy is not viable, then one must rely on the traditional MET, $H_T$, lepton, etc.~triggers, depending on the signature sought.

Because of these complications, it could be tempting to conclude the following:
\begin{loreno}[False]{lore:trigger}
You cannot trigger on LLPs, so you always need to use MET or an associated hard lepton etc.
\end{loreno}
As we will see, this is however not (anymore) the case, as there have been many exciting new developments where dedicated triggers are being developed for LLPs. A nice example is that of displaced muons, for which the muon system can reconstruct muons which do not point to the beamline and/or do not have a matching track in the detector~\cite{ATLAS:2020xyo,CMS:2022qej}. The absence of an inner detector track in particular implies that the momentum resolution for displaced muons at the HLT is substantially worse than for prompt muons.

\begin{figure}
\includegraphics[width=\textwidth]{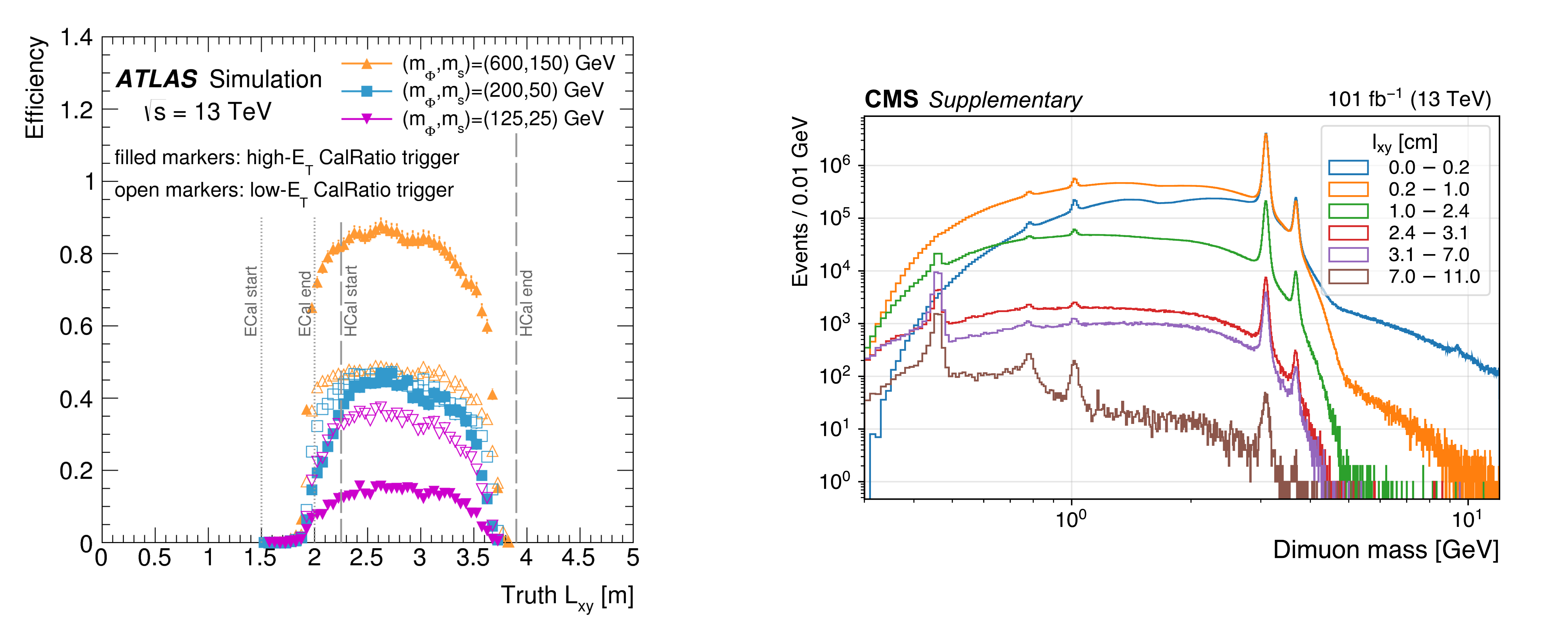}
\caption{\textbf{Left:} Trigger efficiency as a function of the truth-level $L_{xy}$ for the calorimeter ratio trigger at ATLAS \cite{ATLAS:2019qrr}. \textbf{Right:} Inclusive dimuon spectra for multiple $L_{xy}$ bins and $p_T^\mu >  3$ GeV, as measured in the CMS dimuon scouting analysis \cite{CMS:2021sch}. Displaced $\omega/\rho$, $\phi$ and $J/\psi$ mesons can be produced in the decays of boosted $B$ hadrons, hence the persistence of those resonance peaks in the high $L_{xy}$ bins. The $K_S$ resonance peak is due to $K_S\to \pi^+\pi^-$, where both pions were misidentified as muons.
\label{fig:triggers}}
\end{figure}

The ATLAS calorimeter ratio trigger is another successful example \cite{ATLAS:2019qrr}: It takes advantage of the fact that for an LLP decay in the hadronic calorimeter, the ratio of the energy deposited in the ECAL over that in the HCAL is much lower than that for a typical jet. This trigger has good efficiency for the range of $L_{xy}$ that corresponds to the HCAL extent, as is nicely illustrated in the left-hand panel of Fig.~\ref{fig:triggers}.\footnote{Fig.~\ref{fig:triggers} is an excellent example of a ``high value'' plot for theorists seeking to understand or reinterpret an analysis: It presents a very important but detector-specific quantity, the trigger efficiency, in terms of a model-independent, truth-level variable (truth-level $L_{xy}$), which can easily be simulated by theorists using the recipe in Sec~\ref{sec:geoacceptance}. An analogous plot in terms of $c\tau$ would have been much less useful, as one would need to reproduce the exact simulation settings of the collaboration to unfold the $L_{xy}$ dependence of the efficiency, a process which is prone to potential pitfalls.}  Another very interesting option is the possibility to trigger on LLPs decaying in the muon system, mainly because the muon systems have a large fiducial volume, are well shielded by the hadronic calorimeter and are well suited to implement L1 trigger strategies. ATLAS already has such a trigger in place \cite{ATLAS:2013bsk,ATLAS:2022gbw}, while CMS will during Run-3 \cite{Alimena:2021mdu}. A more complete overview of existing and upcoming trigger strategies dedicated to LLPs can be found in the recent LLP working group report \cite{Alimena:2021mdu}.

In some cases it is possible to circumvent the normal bandwidth limitations by only committing a reduced event format to storage, at high rate. CMS and LHCb recently employed this ``data scouting'' technique (called ``Trigger-Level Analysis'' in ATLAS and ``Turbo Stream'' at LHCb) to search for displaced dimuon pairs with very low $p_T$ threshold \cite{CMS:2021sch,LHCb:2019vmc,LHCb:2020ysn}, as shown in the righthand panel of Fig.~\ref{fig:triggers}. 
For Run-3, LHCb moreover eliminates their hardware trigger entirely, such that the data scouting concept can in principle be applied to their entire dataset \cite{TheLHCbCollaboration:2310827}. This is expected to be a major boost to LHCb's sensitivity for low mass LLPs \cite{Borsato:2021aum}. 

In addition to the trigger's computing farms, the experiments also have extensive computing infrastructure which promptly performs the more precise (and more demanding) offline reconstruction for all the events that pass the trigger (see Sec.~\ref{sec:offlinereco}). There are certain cases where a higher trigger rate is desired to be recorded than what this offline reconstruction can keep up with. To achieve this, a fraction of the data can be ``parked'' in raw format until more computing resources are available, e.g.~during LHC downtime. As an interesting example, in 2018 the CMS experiment registered a unique dataset of about $1.2\times10^{10}$ events containing a soft, displaced muon, the signature of a semi-leptonic $B$-hadron decay~\cite{CMS-DP-2019-043}. This data was recorded during the end of LHC fills, when the collision rates and thus the demand on the trigger are lower. Several LLP analyses are ongoing using this dataset, as it is indeed a very interesting opportunity for low-mass LLP searches, by searching for new physics in the decays of the about $10^{10}$ inclusive $B$-hadron decays in this sample.

Unlocking the full potential of the data scouting and parking techniques is still an actively developing and exciting area of research during Run-3, as well as towards the HL-LHC. In general, for the HL-LHC, the detectors will be upgraded with capabilities that will also vastly boost the prospects to trigger on LLPs: Track reconstruction in the Level-1 hardware trigger; ultra-precise timing with dedicated timing layers in front of the calorimeters; impressively improved calorimetry in terms of e.g.~depth segmentation; etc.~\cite{CERN-LHCC-2017-020,Collaboration:2759072}. As was the case for Run-1 and Run-2, we fully expect that the ingenuity of the analysis teams will leverage these new hardware capabilities into often unexpected sensitivity gains for LLPs.

\subsection{Offline signal reconstruction and selection\label{sec:offlinereco}}

Once data events are selected by the trigger and the raw data saved to storage, a more detailed offline event reconstruction takes place, including more precise calibrations and using algorithms with higher complexity and thus more time-consuming than what is possible online. This reconstruction starts from the electronic signals, local hit reconstruction, clustering in higher-level objects such as tracks and jets, and high-level object identification algorithms such as $b$-tagging.
While standard reconstruction is described in detail in numerous references on the detectors and their physics objects, several of the reconstruction steps have interesting features that apply specifically to long-lived particles.

\begin{description}
\item[Displaced tracks:] The arguably most important impact on LLP searches arises from reconstruction of the tracks of displaced charged particles. As mentioned before, optimization of the reconstruction of displaced tracks is a trade-off between efficiency and purity, and in practice it is limited by the available computing resources. The reason is the huge number of hits per event in the inner tracker, which makes track reconstruction a major combinatorical challenge. In the standard track reconstruction, one requires each track to have hits on the innermost layers, which have the best spatial resolution, and to originate from near the beamline. These conditions provide an enormous reduction of the number of possibilities and therefore the complexity of the computational task.

After the prompt tracks are completed, further reconstruction of displaced tracks is attempted with the remaining, unused hits in the tracker. These unused hits are largely from particles produced in nuclear interactions of primary tracks in the detector material, and from particles that were too soft to be considered in the standard reconstruction. For the LHC's current pileup conditions, those unused hits still range in the several thousands per event.
The further reconstruction of displaced tracks with looser constraints, in particular on $d_{xy}$, thus still brings a high level of combinatorial complexity, which in addition scales non-linearly with the amount of pileup. As a consequence, the reconstruction of displaced tracks can only be partially efficient, to avoid picking up too many ``fake'' tracks from nuclear interactions or from combining unrelated hits into a displaced track. Such fake tracks can be a background in certain LLP searches, as is discussed in Section~\ref{sec:fakes}.

Concretely, the inner parts of the ATLAS and CMS detectors are equipped with several layers of high-precision silicon pixel detectors, starting from a transverse distance to the beam of about $3\,\text{cm}$, and yielding typically at least three precise measurements for tracks emerging at a transverse displacement of about $10\,\text{cm}$~\cite{ATLAS:2008xda,CMS:2008xjf}.
Current offline reconstruction algorithms operate at nearly full efficiency for tracks emerging before the first pixel layer -- essential for their superb $B$-hadron identification -- and still show above 50\% efficiency for tracks produced just outside the pixel detector outer radius in LHC Run-2 pileup conditions with $p_{\text{T}} \gtrsim 1\,\text{GeV}$. For larger transverse displacements, an efficiency of about 40\% is still achieved at a track production radius of $30\,\text{cm}$~\cite{CMS:2014pgm,ATLAS:2017zsd}, as shown in Fig.~\ref{fig:displtracking}.
For LHCb, the inner VELO detector~\cite{LHCb:2008vvz} is specifically constructed to be highly efficient for forward displaced charged particles arising in the VELO detector~\cite{LHCb:2014nio}. During Run 3, LHCb will moreover have access to a new class of highly displaced tracks (``T tracks'') \cite{LHCbTtrack}, which will enhance their signal efficiency for LLPs with larger $c\tau$ in particular.

\begin{figure}
\begin{minipage}{.5\textwidth}
  \includegraphics[width=0.95\textwidth]{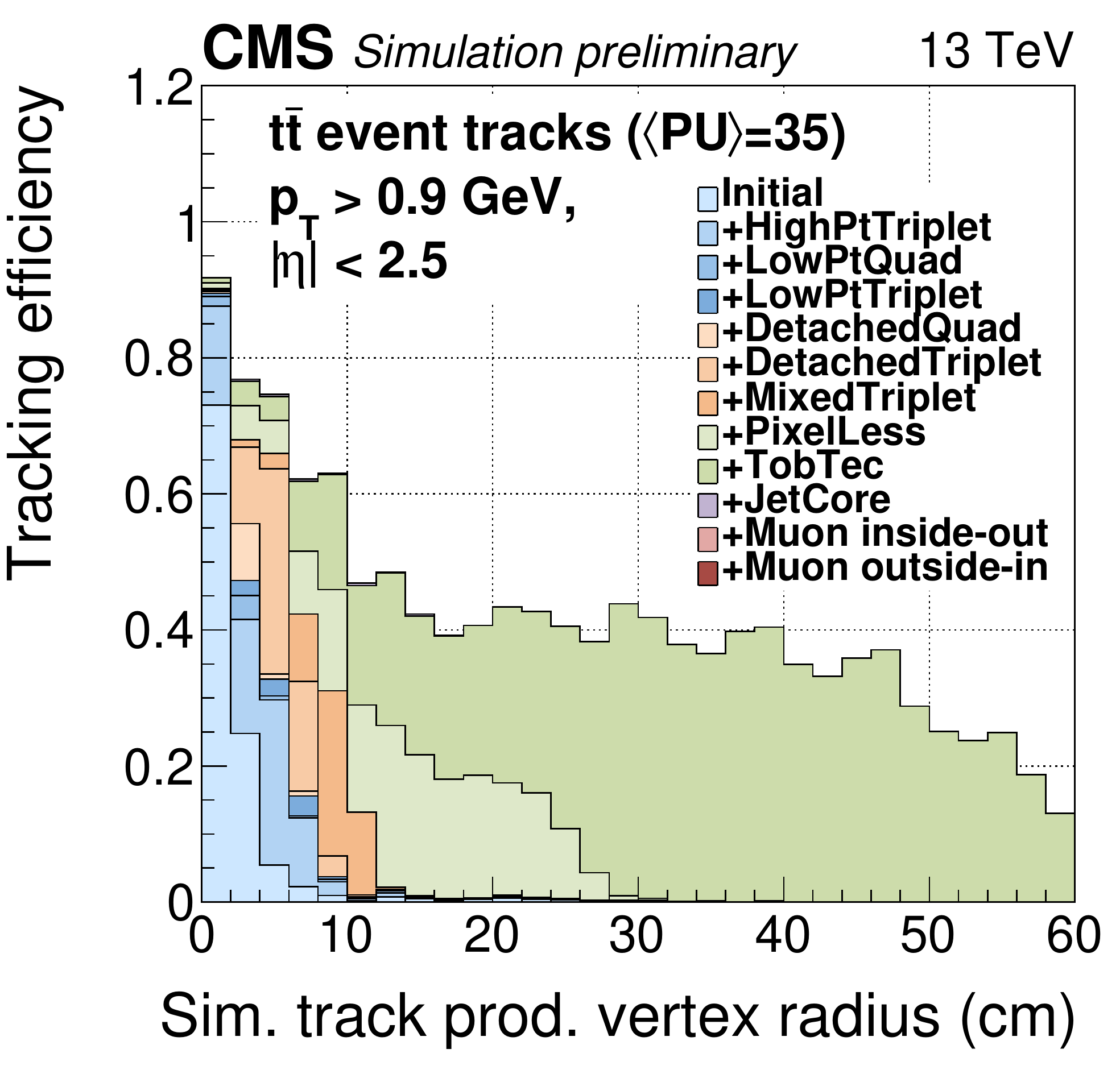}
\end{minipage}%
\begin{minipage}{.5\textwidth}
  \includegraphics[width=0.95\textwidth]{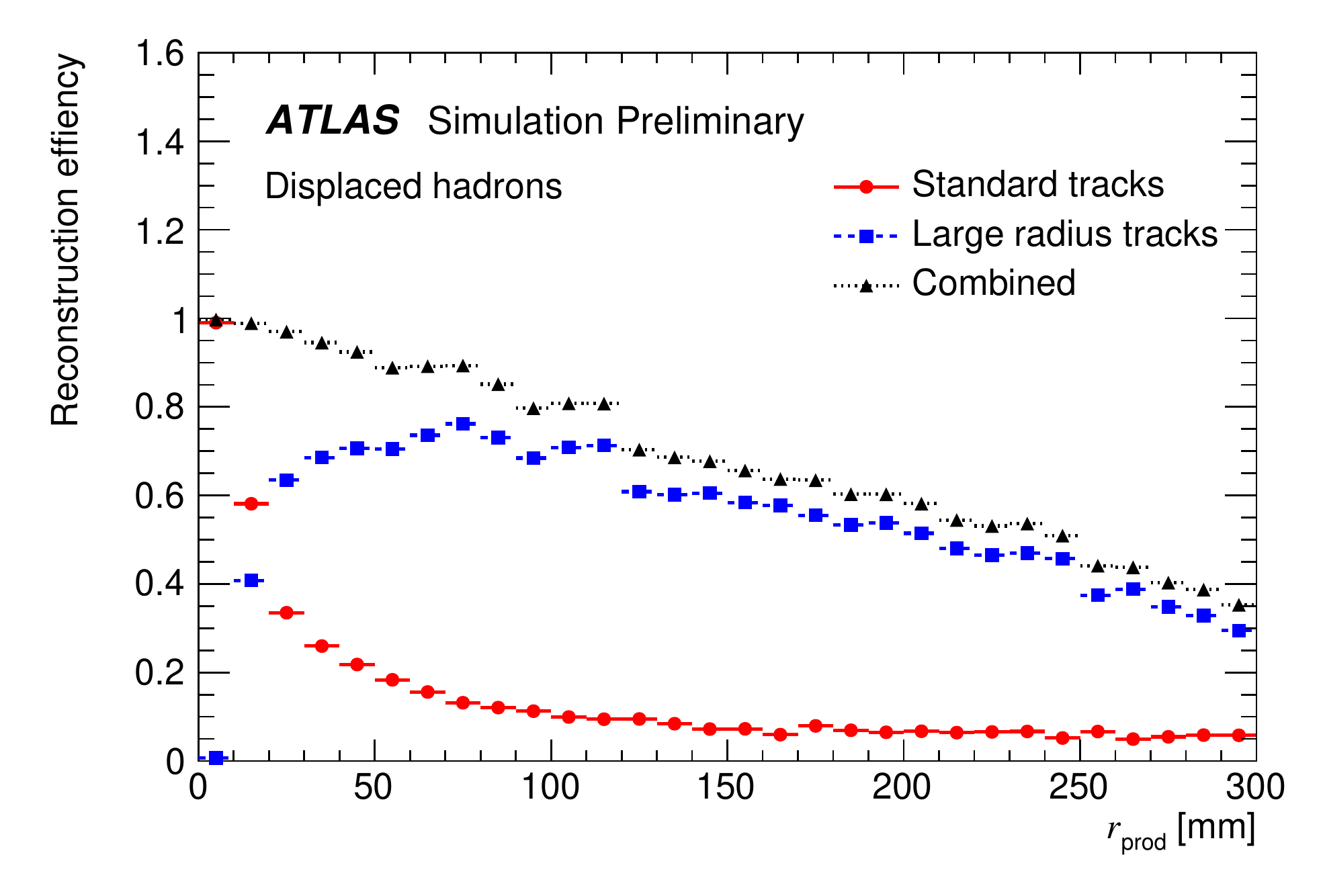}
\end{minipage}
\caption{\textbf{Left:} Track reconstruction efficiency in $t\bar{t}$ events as a function of simulated track production vertex radius (corresponding to $L_{xy}$ in Figure~\ref{fig:trackvars}) for the CMS tracker in its 2017 data-taking configuration. The various colors represent the successive iterations in the reconstruction algorithm, where ever loser criteria are applied to increase the efficiency~\cite{CMS-DP-2017-015}. The green histograms represent tracks reconstructed with algorithms specialized for displaced tracks. \textbf{Right:} Track reconstruction efficiency in ATLAS for displaced charged hadrons produced by the decay of long-lived gluino $R$-hadrons, as a function of the displaced decay radius ($r_\text{prod}$, corresponding to $L_{xy}$ in Figure~\ref{fig:trackvars})~\cite{ATLAS:2017zsd}. The additive efficiencies of the standard and dedicated displaced tracking algorithms are shown.\label{fig:displtracking} }
\end{figure}

The performance calibration of displaced tracking in data is a challenge.
The SM $K_S$ and $\Lambda$ hadrons, commonly referred to as $V^0$'s, are luckily well modeled in simulation~\cite{CMS:2013zgf,ATLAS:2011xhu,LHCb:2011ioc}, and can thus be used as a source of displaced tracks on which displaced tracking performance can be tested.

\item[Short tracks:]  Short tracks are of great interest as an experimental signature of displaced decays of new, charged particles. Often, such particles decay into a nearly mass-degenerate neutral particle that escapes detection, along with a very soft track~\cite{Thomas:1998wy,Gherghetta:1999sw}, or may continue into another charged track with a kinked signature~\cite{Dimopoulos:1996vz}.

In the former case, to reduce backgrounds, a track reconstruction is employed that---unlike standard tracking that is more permissive---requires all consecutive hits on the track to be recorded, with the absence of hits on several outer layers confirming the track has stopped in combination with a quiet calorimetric environment~\cite{ATLAS:2022rme,CMS:2020atg}. The shortest tracks thus reconstructed are promptly produced and have four consecutive hits in the pixel detector. A strong confirmation of such a signal would come from a matching of a soft pion from the short-track's endpoint. However, such a pion would not only appear displaced, it would also circle on a helix in the tracker detector due to its low momentum. This reconstruction is in principle possible, but a costly investment. At this time it is unnecessary, since the background can be suppressed sufficiently in other ways.

Regarding kinked tracks, there is currently no dedicated analysis for this signature. That said, the existing disappearing track searches likely have excellent sensitivity already~\cite{Evans:2016zau}. This is because the outer part of the kinked track typically fails to be reconstructed, such that the signature is effectively identical to that of a disappearing track.

\item[Displaced muons:] For muons with a matched inner tracker track, the efficiency is set by the inner track reconstruction, as described above. Muons can also be reconstructed up to very large displacements beyond the inner tracker using the muon system only~\cite{CMS:2022qej,ATLAS:2018rjc}, albeit with a lower position and momentum resolution. This can be done very efficiently up to track transverse displacements as large as $400 \, \mathrm{cm}$, imposing quality requirements that suppress backgrounds from hadrons punching through the calorimeters and other non-beam backgrounds (see Section~\ref{sec:background}). For these muons, it is important to not use the collision vertex as a constraint, as it biases the reconstructed momentum to lower values, and induces inefficiencies at large impact parameter values. 

\item[Displaced electrons, taus and photons:] For displaced electrons originating in the first part of the tracker, the efficiency will mostly mirror the displaced track reconstruction performance. The reconstruction efficiency of such displaced electrons can be estimated in data by looking at photon conversions in the detector material~\cite{CMS:2022fut}. If its track is too displaced to be reconstructed but emerges before the EM calorimeter, the electron can still be reconstructed as a photon. While this comes at the cost of a higher background, it is particularly useful in the trigger~\cite{CMS:2021kdm,ATLAS:2020wjh} prior to a more detailed offline selection.

Genuine photons from LLP decays benefit from their own dedicated treatment, with adapted identification requirements on timing or direction~\cite{ATLAS:2022vhr,CMS:2019zxa}. Also tau leptons emerging from displaced decays are an interesting target, either as a displaced electron or muon, or a displaced jet-like signature. Dedicated displaced hadronic tau identification is the most complex final state, which we expect to see develop strongly during LHC Run-3.

\item[Displaced vertices:] Once the displaced tracks have been reconstructed, the reconstruction of their corresponding displaced vertex is virtually 100\% efficient. Nevertheless, in cases where the tracking efficiency is reasonably inefficient, it may still be better to forgo vertex reconstruction, as the overall efficiency will scale as the tracking efficiency raised to the number of tracks required to reconstruct the vertex. In particular, if one expects multiple displaced decays in the event, it may therefore be more beneficial to ask for a number of displaced tracks, without insisting that they belong to a single vertex \cite{CMS:2018bvr}.

\item[Exotic objects:] In the above, we discussed aspects of offline reconstruction of the most common final state objects. More specialized reconstruction furthermore targets specific experimental signatures, some of which, like showers in the muon system, were touched upon in Section~\ref{sec:otherexphandles}. The potential for future development of very advanced tracking for exotic tracks from monopoles is also exciting, as they are bent along the magnetic field direction. Quirks oscillating in pairs when traversing the detector~\cite{Kang:2008ea,Knapen:2017kly} are another example.

\end{description}

A recent avenue of significant progress is the development of more advanced techniques which aggregate various sources of experimental input into a multivariate discriminator. This can significantly boost the sensitivity of a search~\cite{CMS:2019dqq,ATLAS:2022zhj}. The potential downside of less detailed control of a selection cut can be offset by using the neural network as a discriminator to select physics objects, coupled with more robust background predictions from the data. The simulation description of the multivariate discriminator, on the other hand, is difficult to assess because of the limited types of control samples with genuine displaced tracks available. Here, techniques like domain adaptation~\cite{CMS:2019dqq} and adversarial networks~\cite{ATLAS:2022zhj} alleviate this challenge.

As ever more advanced novel reconstruction algorithms are developed, also lower-level detector features come in focus. This is at odds with the increasing computing complexity, which requires one to use as high-level and thus compact objects as possible. Advanced offline reconstruction thus benefits from being integrated in the standard reconstruction chain (see e.g. Ref.~\cite{ATL-PHYS-PUB-2021-012} for the integration of ATLAS Large Radius Tracking into the main reconstruction from Run-3 onwards), or must run on dedicated data streams where low-level objects are being kept availabe for such dedicated analysis purposes. Either way, such efforts require sometimes heroic, long-term investments.

\section{Backgrounds\label{sec:background}}

Searches for LLPs make use of non-standard experimental signatures in both online and offline selection to bring the prompt backgrounds down dramatically at a modest cost in signal efficiency. This has led to the following, general assumption in many LLP theory studies
\begin{loreno}[False]{lore:backgroundfree}
LLP searches are background free.
\end{loreno}
Though in many analyses the backgrounds can indeed be reduced to be negligible, this is only possible after extensive and subtle analysis efforts on the experimental side. Even after such extensive background reduction efforts, often unusual irreducible backgrounds from instrumental, algorithmic, or other origins remain. In the following, we review various sources of backgrounds in searches for LLPs and some methods employed to eliminate them. Often, simulations are not reliable or robust enough to estimate the remaining background, even if very small, and data-driven techniques are needed to quantify them reliably.

\subsection{Standard Model long-lived particles\label{sec:realbackground}}

Charm and bottom flavored hadrons are the most ubiquitous SM background for LLP searches, as they can easily produce a multi-track displaced vertex. A priori they can be removed effectively with a cut on the vertex mass of $\gtrsim 5$ GeV (see e.g.~\cite{LHCb:2020akw}), though this severely limits the sensitivity to low mass LLPs. Their proper decay length is also only $\mathcal{O}(0.1)$ mm, such that cuts on $d_{xy}$ and/or $L_{xy}$ are also very effective, see e.g.~\cite{CMS:2021sch,LHCb:2017xxn} and the right-hand panel of Fig.~\ref{fig:triggers}. One must be aware however that there are still many events with highly boosted $b$ or $c$-jets, which could still leak into the signal region in some cases. When using simulation to estimate these backgrounds, it is therefore essential to use $p_T$-weighted events. Whenever possible it is moreover advisable to also reweight the events in terms of the long-lived meson's decay vertex location, as described in Sec.~\ref{sec:geoacceptance}.

The charged pions have a proper decay length of 780 m, and the vast majority of them therefore reaches the calorimeter without decaying. Bearing in mind Sec.~\ref{sec:geoacceptance} however, we can ballpark the probability for a $\pi^\pm$ to decay in the tracker. If we suppose a $\pi^\pm$ with $\beta\gamma \sim 10$, we find $1\,\text{m} /(10 \times 780\, \text{m})\approx 10^{-4}$. This seems very small, but we must keep in mind that \emph{every} collision produces tens to hundreds of $\pi^{\pm}$, though most of them are soft. Multiplying this with the $\mathcal{O}(100)$ for the number of pileup collisions per event, we see that an $\mathcal{O}(1)$ fraction of all events will have a $\pi^{\pm}$ which decayed ``early''. Similar considerations apply for the $K^\pm$, $K_L$ and especially the $K_S$. Both pions and charged kaons are therefore a huge potential source of individual displaced muons in particular. Fortunately, this background drops quite rapidly if isolation is imposed and if the $p_T$ cut on the muon is tightened. Kaons can moreover produce a genuine displaced vertex. For example, the right-hand panel of Fig.~\ref{fig:triggers} shows a clear peak in the kaon mass range, which is due to $K_S\to \pi^+\pi^-$, where both pions are misidentified as muons. In general such displaced vertices can easily be eliminated by requiring vertex mass well above the kaon mass and/or demand a larger number of tracks.

Jets with an anomalously low number of tracks are an interesting handle for displaced decays near the back of the tracker or in the calorimeter \cite{ATLAS:2022zhj} or for strongly interacting dark matter candidates \cite{CMS:2021rwb}. A priori, QCD can produce jets primarily with $K_L$ and neutrons, which don't leave tracks but do deposit energy in the HCAL. While the corresponding probability per jet is very low, this background can nevertheless be important due to the huge QCD multi-jet cross section. 

For all the above backgrounds, the most important point is perhaps to be very mindful of the limitations of all simulation codes, as they are not designed to model very rare effects in very specific corners of phase space. This can to some extent be offset with the weighting procedures described in Sec.~\ref{sec:geoacceptance}, but more often than not a data-driven validation is needed.

\subsection{Material interactions}

\begin{figure}
\begin{minipage}{.37\textwidth}
  \includegraphics[width=0.95\textwidth]{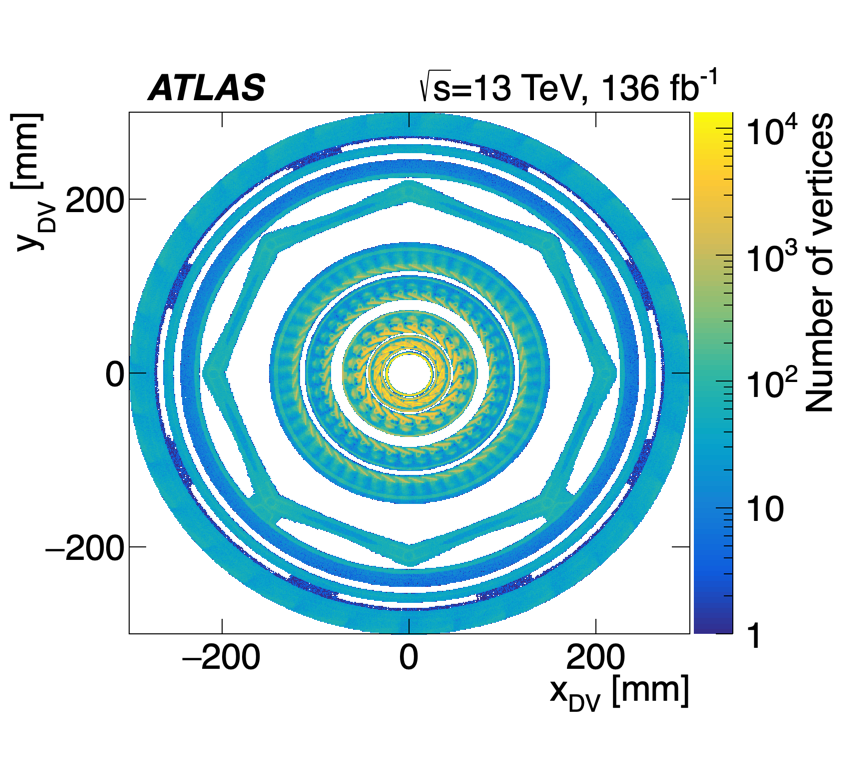}
\end{minipage}%
\hfill
\begin{minipage}{.57\textwidth}
  \includegraphics[width=0.95\textwidth]{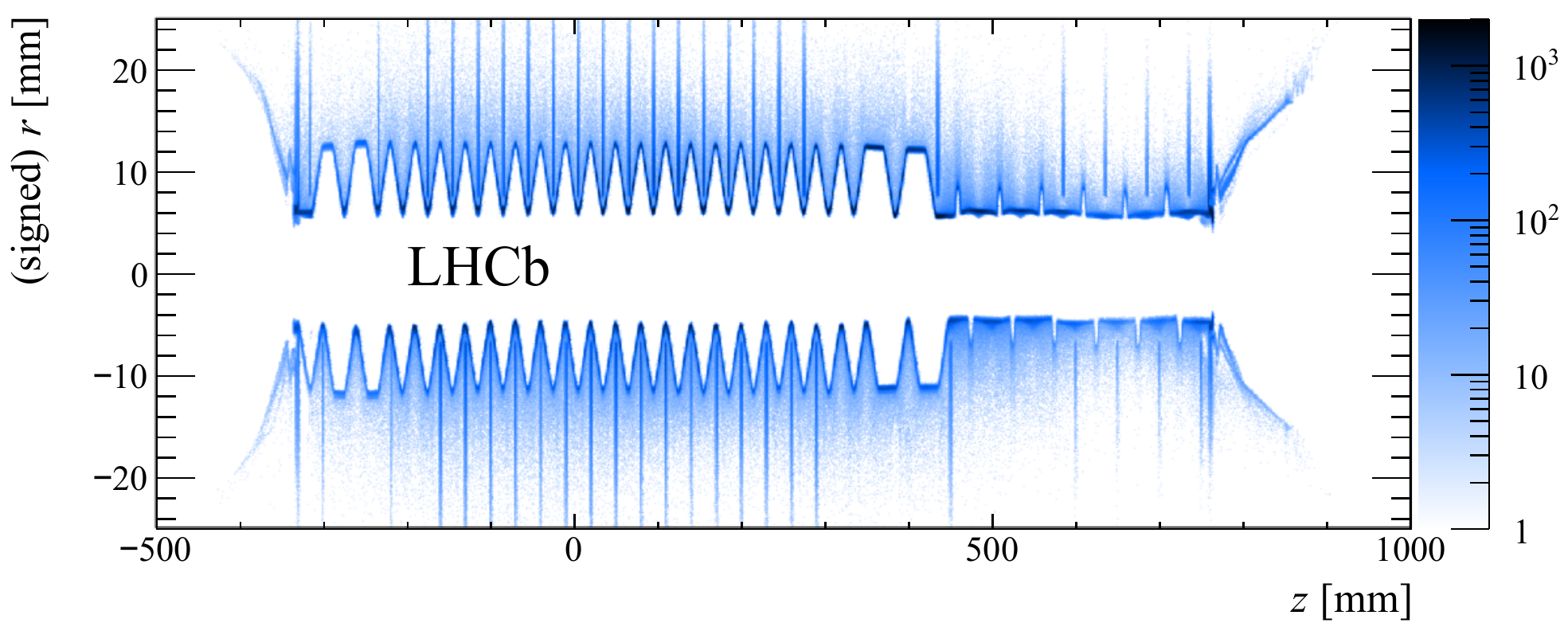}
\end{minipage}%
\caption{\textbf{Left:} Number of reconstructed secondary vertices in the ATLAS inner tracker \cite{ATLAS:2022fag,ATLAS:2017tny}. \textbf{Right:} Same, for the LHCb Vertex Locator (VELO) \cite{Alexander:2309964}. \label{fig:materialinteractions}}
\end{figure}

The details of the detector material, its density and geometry are critical for LLP searches, as it both reduces and generates more backgrounds. First, the trackers are always designed to have as a little material as possible, as particles scattering in sensors or support structures adversely affect the momentum resolution of the tracks. This design driver is also helpful for LLP searches, as inelastic collisions or photon conversions in detector material can produce secondary vertices. This is \emph{not} a rare effect by any means: About 5\% of all $\pi^\pm$ with $p_T\gtrsim 5$ GeV create a secondary vertex within the CMS inner tracker \cite{CMS-PAS-TRK-10-003}. Extrapolated to HL-LHC conditions, this implies secondary vertices at a rate of $\sim 30$ MHz, essentially in every event! Fortunately, this background is very strongly dependent on the track $p_T$, the vertex mass and the number of tracks per vertex  \cite{CMS-PAS-TRK-10-003}. The exquisite vertex resolutions of their trackers moreover allow ATLAS, CMS and LHCb to make beautiful maps such as Fig.~\ref{fig:materialinteractions}, pinpointing precisely where the vertices are produced. This can then be interpreted as a radiography of the detector material and be used as a veto map to suppresses backgrounds extremely efficiently in a real LLP analysis. To avoid masking too much of the detector volume, which reduces signal efficiency, further track and vertex selections can be made to improve the vertex resolution, e.g.~avoiding vertices from very collimated tracks. Kinematic cuts may also prove useful against SM LLPs produced in material interactions, which may decay in the unmasked region.

Unlike the trackers, the calorimeters are designed to stop as many particles as possible. Especially the hadronic calorimeters are made out of heavy elements and are made as thick as the engineering constraints allow. For example, the ATLAS HCAL in the barrel is made up of $\sim$ 2 m of steel plus scintillator tiles. At $\eta\approx0$, this corresponds to roughly 7.4 nuclear interaction lengths or 9.7 nuclear interaction lengths when we include the liquid argon ECAL. We can therefore estimate the probability for a hard $\pi^\pm$ to punch through the calorimeter into the muon system to be $e^{-9.7}\approx 5\times 10^{-5}$. %
This is small, but not very small, given the huge number of  $\pi^\pm$ that hit the calorimeter during a typical run. This means that large numbers of hadrons still punch through the calorimeters into the muon chambers, where they may fake a muon or a displaced vertex. Fortunately, hard hadrons typically live inside a jet, and the calorimeter itself is therefore a very effective veto for such ``punch-through'' events. This background cannot be simulated faithfully and must be estimated from data. A search along these lines was first performed by ATLAS, using a dedicated LLP trigger on activity in the muon system \cite{ATLAS:2015xit,ATLAS:2022gbw}. The CMS calorimeter is significantly thinner, and thus a priori suffers from more punch-through backgrounds. CMS can however offset this by using some of the steel in the return yoke of its magnet as an additional shield \cite{CMS:2021juv}.

\subsection{Fake tracks and vertices\label{sec:fakes}}

While material interactions inject real displaced particles in the detector, tracks may also be reconstructed that do not arise from a real particle. Fake hits from detector noise, or more likely hits from the cloud of thousands that come from the pileup particles in the inner detectors may align up to form a track. For example, assuming on average 50 pileup vertices, the four innermost tracking layers in the ATLAS barrel collect on average about $\sim 2\times 10^4$ hits per bunch crossing \cite{ATLASpixel}.  Roughly 25\% of these hits will be associated with reconstructed, prompt tracks, leaving several thousands of unassociated hits in each event. While it is very unlikely that such unrelated hits will accidentally line up and have a good fit quality for the hypothesis of prompt production, this becomes more likely for displaced tracks and when only a handful of hits are required, as e.g.~in searches for disappearing tracks~\cite{ATLAS:2022rme,CMS:2020atg}, or in cases of displaced muon reconstruction~\cite{ATLAS:2018rjc,CMS:2022qej}.

Fake vertices are also common, and can be composed out of a set of real but unrelated tracks. This frequently happens when a vertex from muons with very large displacements is reconstructed, or vertices with low displaced track multiplicity are considered in the inner detector. A particular source of fakes arises from overlapping tracks from different pileup vertices. While in this particular case, such fakes can be identified from the tracks lining up in a plane with the beamline, more generally fake vertices may be rejected by vetoing hits on the tracks upstream of the vertex.

Backgrounds from fake tracks and vertices are best measured from data, though there have been attempts to estimate them in phenomenological studies \cite{Gershtein:2017tsv,Gershtein:2019dhy,Hook:2019qoh}. In general they must however be extracted from data, and caution is needed especially when a very high degree of background rejection is required.

\subsection{External particle backgrounds}

The above sources of background relate directly or indirectly to the proton collisions. There are however a few additional sources of background that may enter the signal phase space of long-lived particle searches.

\begin{description}

\item[Beam halo] As the protons travel through the LHC in dense bunches, some stray too far from their ideal trajectory, posing problems to the sensitive equipment of the LHC and the experiments. Dedicated absorbers clean the proton beam of such beam halo, during which beam halo muons may be produced that travel along with the proton bunch parallel to the beam and may traverse the thick shielding in front of the detectors unhampered. The rate of such muons decreases strongly as a function of the radial distance to the beam  \cite{Drollinger:883295}.

These muons may create some harmless background hits in the tracking detectors, but in the calorimeters and muon spectrometer they can lead to unusual backgrounds. Since the muons do not originate from the proton collisions, their timing is asynchronous, though in a predictable manner. They may leave straight muon tracks in the forward muon tracker, as well as significant energy deposits in the calorimeters. The parallel direction of the muon tracks, the potential match of these tracks with calorimeter deposits, the anomalous shapes of such deposits parallel to the beamline, depth information, and the early or late timing of these signatures may be used to suppress this background. The azimuthal $\phi$ distribution is another interesting handle: The beam halo rate spikes in the horizontal plane and is smallest near the bottom of the detector, as the floor of the LHC tunnel acts as a shield \cite{Drollinger:883295,CMS:2017qyo}.

\item[Cosmic muons] A fraction of the relativistic muons that are created in cosmic ray showers reach underground and cross the detectors downward with a rate of roughly $1\,\mathrm{Hz/m^2}$ \cite{Drollinger:883295}. Their energy spectrum falls exponentially and they exhibit a significant spread in direction and a geometric asymmetry, mainly due to the access shafts above the detectors.  Only a small fraction is reconstructed as a track or a muon, since the cosmic muon only rarely has a direction that is compatible with the constraints of the track reconstruction algorithms. In addition, the arrival time of the muon is random, and thus the timing of the energy deposits in the various subdetectors, including calorimeters, is not syncronized with the collision it may have an overlap with. For downgoing tracks in the upper half of the detector, the travel direction is even opposite of what is expected for particles originating from collisions. Given these features of cosmic muons, they are only a potential background in LLP searches, where anomalous timing or track displacement are selected, see e.g.~\cite{CMS:2022qej,ATLAS:2022izj}.

\item[Satellite collisions] Though the LHC beam delivers its main bunches in 25 ns intervals, the beam inevitably also has ``satellite bunches'', which follow and precede each main bunch with a 2.5 ns time gap. These satellite bunches contain about $10^{-5}$ times the number of protons of the main bunches and their collisions can generate very rare, out-of-time backgrounds. Such backgrounds arrive with well-defined $5\,\mathrm{ns}$ delays as compared to the primary collision and can be important particularly for analyses focusing on delayed calorimeter signals, e.g.~\cite{CMS:2019qjk}.

\end{description}

\section{Dedicated detectors\label{sec:dedicated}}

As we have seen, ATLAS, CMS and LHCb are excellent detectors to hunt for long-lived particles. There are however a couple important cases where dedicated long-lived particle detectors can perform qualitatively better. Concretely, there are three reasons to consider auxiliary detectors at relatively large distances from the interaction point: \emph{(1)} to catch LLPs with very forward kinematics, which are inaccessible to the main detectors, \emph{(2)} to allow for a different detector technology and \emph{(3)} to allow room for additional shielding. All proposals rely on \emph{(3)} to some extent.

The FASER experiment  \cite{Feng:2017uoz} makes use of a small service tunnel, 480 m forward of the ATLAS interaction point. Its extreme forward location allows it to search for LLPs produced through the huge pion flux that goes down the beam pipe with each collision. A larger version of FASER would need to be housed in a proposed, dedicated cavern, the forward physics facility (FPF)~\cite{Feng:2022inv}. The FACET experiment \cite{Cerci:2021nlb} would follow a similar philosophy and would be located roughly 100 m forward from CMS. It would cover a somewhat larger pseudo-rapidity range than FASER, though beam backgrounds may be more challenging to mitigate. Finally, the SND$@$LHC detector \cite{SHiP:2020sos} is meant to detect high energy, forward neutrinos from the ATLAS interaction point, but may also have sensitivity to some low mass LLPs \cite{Boyarsky:2021moj}. When modeling the acceptance of forward LLP detectors, it is important to remember that standard simulation tools such as PYTHIA and MadGraph are typically not appropriate in this kinematical regime. Instead specialized tools are needed, see e.g.~\cite{Pierog:2013ria}.

The MoEDAL~\cite{MoEDAL:2014ttp} and milliQan~\cite{Haas:2014dda} experiments rely on alternative detector designs, specialized to stable LLPs which leave anomalous tracks. MoEDAL has searched for magnetic monopoles which get trapped in a set of aluminum rods, which were scanned for monopoles with a SQUID sensor \cite{MoEDAL:2017vhz}. The MoEDAL collaboration also aims to install an extension which can look for decaying LLPs at intermediate rapidities \cite{Acharya:2022nik}. Similar to FASER and SND$@$LHC, the milliQan experiment is also housed in an existing service tunnel, but at moderate rapidity above CMS. It is shielded by 17m of rock and designed to detect fractionally charged particles. It can do so by looking for coincident hits in 4 aligned bars of plastic scintillator. 
Currently, a first modular phase of the detector is under construction~\cite{milliQan:2021lne}, with sensitivity opening up a large new phase space during LHC Run-3 and HL-LHC.

Finally, proposed experiments such as CODEX-b \cite{Gligorov:2017nwh,Aielli:2019ivi} and MATHUSLA \cite{Chou:2016lxi,MATHUSLA:2018bqv} would look for displaced vertices in a similar manner as the ATLAS and CMS muon chambers, but would come with a much thicker shield between the detector and interaction point. This results in much lower backgrounds, at a cost in geometric acceptance. MATHUSLA would be constructed on the surface above CMS, using the rock as shielding. CODEX-b would be installed directly in the LHCb cavern and a suitable shield would need to be constructed. In either case the shield must be equipped with an active muon veto to confidently reject muon-induced secondaries as potential backgrounds \cite{Gligorov:2017nwh,Aielli:2019ivi,Gligorov:2018vkc}.

\section{Closing thoughts}

The ability to reconstruct and characterize displaced particle decays has historically delivered many discoveries and continues to be an essential tool for particle identification today. It is moreover a misconception that heavy, beyond the SM particles are supposed to decay promptly: Not only do there exist many counterexamples, the general conditions for macroscopic lifetimes are also very simple and likely generic, given our experience with the SM itself. Searches for displaced signatures moreover benefit from a number of experimental handles which are unavailable for promptly decaying new particles and can therefore often maintain very low backgrounds, even with very high integrated luminosity. Long-lived particles are therefore one of the primary areas where we may achieve a major discovery at HL-LHC.

Searches for long-lived particles are also fascinating because they are complicated, subtle and often messy. The signal efficiency depends on the details of the detector design, trigger capabilities and limitations, as well as the available offline reconstruction methods. The backgrounds are delicate and sometimes impossible to simulate; understanding them requires careful and clever estimation techniques. The detector capabilities moreover continue to evolve, especially with the upcoming HL-LHC upgrades, all while new trigger strategies are constantly added, unlocking qualitatively new searches. All of this means that there is still a lot of room for new ideas which may impact the direction of the ongoing and future search programs.

The flip-side is that the up-front investment can seem very high for those seeking to get started in this field, especially as a good deal of the essential knowledge is often either unwritten or buried deep in fist-thick technical design reports. In this review we have attempted to collect some of this knowledge in a manner that aims to be accessible to novices, along with some practical suggestions. We conclude with a few further words of advice for both beginning theorists and experimentalists. In particular for theorists, 
\begin{itemize}
\item It pays off to learn how to read technical design reports and performance papers. While they may seem daunting at first, many are structured in a similar manner, and after a while you would be surprised how quickly you can mine them for the trigger and/or reconstruction efficiencies you need.
\item Be realistic when it comes to your ability to accurately model complicated backgrounds. Rather than producing overly aggressive or overly conservative (projected) limits, consider plotting signal yields only, and let experimentalists follow up with a full analysis.
\item Consult extensively with experimental colleagues, either by collaborating directly or by reaching out to the authors of the analysis you are studying. Figures may rely on important assumptions which may not be obvious from the published material. The conveners of the analysis groups can put you in touch with the main analysis authors. 
\end{itemize}
For experimentalists, we suggest to keep the following in mind:
\begin{itemize}
\item Connect with the experts in the individual subdetectors, who are often eager to understand together low-level issues in the detector data or simulation. Also share your challenges and progress regularly with your colleagues, as similar problems may find solutions in very different contexts.
\item Plan ahead any special trigger or computing needs, including data storage and access. Also prepare to release publicly efficiencies, cut-flow tables, etc, to allow theorists to recast your analysis, which greatly amplifies its impact (see e.g.~\cite{CMS:2021sch}).
\item Keep abreast of the phenomenology literature and community. Theorists are generally happy to connect and follow up on their work. This may even lead to fruitful common research.
\end{itemize}

Long-lived particles provide a compelling window on physics beyond the SM, and a rich research arena for both experimentalists and theorists. With this review, we hope to have conveyed this excitement, with links, tools and advice that may help new physicists make the leap towards long-lived particles.


\section*{DISCLOSURE STATEMENT}
The authors are not aware of any affiliations, memberships, funding, or financial holdings that
might be perceived as affecting the objectivity of this review. 

\section*{ACKNOWLEDGMENTS}

We would like to sincerely thank Dean Robinson, Simone Pagan Griso, Jessie Shelton, Matthew Citron, Daniele Del Re and Brian Shuve for their very valuable inputs to this review. We are particularly grateful to Jessie Shelton, Federico Leo Redi and Alberto Escalante del Valle for providing comments on the manuscript.
The work of SK was supported by the U.S. Department of Energy, Office of Science under contract DE-AC02-05CH11231.
Part of this work was performed at the Aspen Center for Physics, which is supported by National Science Foundation grant PHY-1607611.
SL would like to acknowledge 2021 sabbatical leave support from Vrije Universiteit Brussel and support of FWO Vlaanderen under the “Excellence of Science – EOS” – be.h project n. 3082081.

%
\bibliography{LLP_review}
 
\end{document}